\documentclass[aps,prl,twocolumn,epsfig,pdflatex]{revtex4-1}

\usepackage{amsmath,amssymb}
\usepackage{graphicx}

\usepackage{float}
\usepackage[usenames,dvipsnames]{xcolor}
\usepackage{amsthm,comment}
\usepackage{booktabs}
\usepackage[export]{adjustbox}
\usepackage[section]{placeins}
\usepackage[caption=false]{subfig}
\usepackage[percent]{overpic}
\bibpunct{[}{]}{,}{n}{}{}
\definecolor{red1}{HTML}{FF4136}
\definecolor{green1}{HTML}{00802b}

\usepackage[hidelinks]{hyperref} 

\hypersetup{colorlinks=true,citecolor=Red,linkcolor=Blue,urlcolor=Blue}
\usepackage[export]{adjustbox}

\graphicspath{{Figures/}}

\def\ups{\uparrow}
\def\downs{\downarrow}

\begin{document}

\title {Intertwined charge, spin, and pairing orders in doped iron ladders} 
\author{Bradraj Pandey$^{1,2}$, Rahul Soni$^{1,2}$, Ling-Fang Lin$^{1}$, 
Gonzalo Alvarez$^{3}$, and Elbio Dagotto$^{1,2}$}
\affiliation{$^1$Department of Physics and Astronomy, University of Tennessee, Knoxville, Tennessee 37996, USA \\ 
$^2$Materials Science and Technology Division, Oak Ridge National Laboratory, Oak Ridge, Tennessee 37831, USA \\ 
$^3$Computational Sciences $\&$ Engineering Division and Center for Nanophase Materials Sciences,
Oak Ridge National Laboratory,~Oak Ridge,~Tennessee 37831,~USA
 }

\begin{abstract}

        Motivated by recent experimental progress on iron-based ladder compounds, we study the doped two-orbital Hubbard 
	model for the two-leg ladder BaFe$_2$S$_3$. The model is constructed by using {\it ab initio} hopping parameters and the ground state 
	properties are investigated using the density matrix renormalization group method. We show that the $(\pi,0)$ magnetic ordering at half 
	filling, with ferromagnetic rungs and antiferromagnetic legs, becomes incommensurate upon hole doping. Moreover, depending on the strength 
        of the Hubbard $U$ coupling, other magnetic patterns, such as $(0,\pi)$, are also stabilized. We found that the
	binding energy for two holes becomes negative for intermediate Hubbard interaction
	strength, indicating hole pairing. Due to the crystal-field split among orbitals, the holes primarily reside in one orbital, 
        with the other one remaining half-filled. This resembles orbital selective Mott states. 
        The formation of tight hole pairs continues with increasing hole density, as long as the magnetic order remains 
        antiferromagnetic in one direction. The study of pair-pair correlations indicates the dominance of the intra-orbital 
        spin-singlet channel, as opposed to other pairing channels. Although in a range of hole doping pairing correlations decay slowly, 
        our results can also be interpreted as corresponding to a charge-density-wave made of pairs, a precursor of eventual 
        superconductivity after interladder couplings are included. Such scenario of intertwined orders has been extensively 
        discussed before in the cuprates, and our results suggest a similar physics could exist in ladder iron-based superconductors.
        Finally, we also show that a robust Hund's coupling is needed for pairing to occur.

\end{abstract}


\maketitle

\section{I. Introduction}
The study of iron-based high critical temperature
superconductors continues attracting considerable attention in the Condensed Matter community~\cite{kami,wat,taka,chen,luet,fang,chi,john,stewart,guo,fang1} .
The parent compounds of iron-based superconductors exhibit non-trivial magnetic ordering~\cite{bao,elbio,dong} and 
can have either metallic or insulating characteristics
~\cite{john,stewart,guo,fang1}. 
Early theoretical studies based on weak-coupling theory
suggested that antiferromagnetic (AFM) order is stabilized by Fermi surface
nesting~\cite{mazin,korsh}, leading to the prediction of 
superconductivity with $s^{\pm}$ paring symmetry induced by AFM fluctuations~\cite{pagl,kuroki}.
However, the absence of hole pockets in some compounds~\cite{liu,guo} 
and the presence of robust local magnetic moments~\cite{bondi} indicated that the role of 
electronic correlations can not be neglected~\cite{dai,dong,elbio,fang1,wei,ying}. 
Superconductivity can be induced in iron-based compounds by either electron or hole 
doping~\cite{stewart,pagl} or also by applying pressure~\cite{ueda,nambu}.
For example, the iron-based compound BaFe$_2$As$_2$ becomes superconducting by 
electron doping, namely by partially replacing Fe by Ni~\cite{chen}.
Also superconductivity has been induced in the hole doped
Ba$_{1-x}$K$_{x}$Fe$_2$As$_2$~\cite{avci} and it can even survive the hole overdoped regime~\cite{lee}. 
Interestingly, by hole doping BaFe$_2$As$_2$, the 
superconducting  transition temperature $T_c$ can reach higher values than by electron doping~\cite{rotter}.

More interesting for our present study, 
pressure-induced superconductivity has been achieved in geometries different from planes, namely 
in the iron-based ladder material BaFe$_2$S$_3$~\cite{ueda,nambu}.
In particular, 
BaFe$_2$S$_3$ becomes superconducting at pressures above  $10$ GPa
with critical temperature $T_c=24$ K~\cite{ueda}. 
At ambient pressure this material
is a Mott insulator with stripe-type arrangement, where
magnetic moments align ferromagnetically along the ladder rung direction and
coupled anti-ferromagnetically along the legs of the ladder~\cite{nambu}.
This exciting experimental progress in quasi one-dimensional iron-based compounds
 provides a promising platform to explore the magnetism and 
 superconductivity in iron-based materials~\cite{caron,sipos,llobet,hirata,
 stone,ueda,nambu,ling,Yzhang,nitin} particularly from a theoretical perspective.
In particular, due to the availability of powerful numerical many-body techniques for 
 quasi-one dimensional systems, it is possible to address fairly accurately 
 the ground state properties of iron-based ladder compounds using complex 
two-orbital Hubbard models incorporating
quantum fluctuations, without resorting to crude many-body approximations~\cite{luo,patel, pandey,arita}.

Already exotic theoretical predictions for iron ladders have been confirmed experimentally.
For example, using inelastic neutron diffraction applied to powder
BaFe$_2$Se$_3$, an exotic block-AFM state (involving blocks of 2$\times$2 iron atoms 
ferromagnetically aligned, coupled anti-ferromagnetically along the leg direction) 
was observed after theory predicted such a state~\cite{caron,llobet,ryu,sato}. 
The compound BaFe$_2$Se$_3$ shows insulating behavior, 
with energy gap  $\Delta \sim 0.13$~eV to $0.178$~eV~\cite{ryu,sato}. This 
compound displays long-range antiferromagnetic order at $\sim 250$~K, presumably from interladder coupling,
with large individual magnetic moments $2.8 \mu_B$~\cite{caron,sipos,ryu}. 
Interestingly another compound where K replaces Ba, i.e. KF$_2$Se$_3$, 
shows a magnetic arrangement with ferromagnetic rungs coupled 
anti-ferromagnetically along the legs~\cite{miller}.

 Because the work in the related field of computational studies of superconducting Cu-oxides ladders 
was extremely useful in showing
that pairing can emerge from repulsive interactions~\cite{riera,rice,ueda,nagata,uchida}, this provides additional
 motivation to proceed with the numerical studies of 
iron-based ladder models as well.
Experimental work has shown that both Fe- and Cu-based compounds 
induce superconductivity via a magnetic
pairing mechanism~\cite{elbio,dai,pagl}.
However, technically there is a practical difference. While Cu-oxide ladders can be described by 
a one-orbital Hubbard $U$ model~\cite{dagotto}, the Fe-based ladders require 
a multiorbital Hubbard $U$ description that must also include the
Hund coupling~\cite{dai}. Increasing the number of orbitals quickly increases the
computational effort.
However, the competition between charge, 
spin, and orbital degrees of freedom in Fe-based compounds 
can also lead to various exotic novel phases
with insulating or metallic ground states~\cite{elbio,dai}.

In the present publication, we study the magnetic and pairing
properties of hole-doped two-orbital Hubbard model using the
density matrix renormalization group method (DMRG)~\cite{white}.
For our DMRG calculations, we use realistic hopping parameters 
for the compound BaFe$_2$S$_3$, originally derived in Ref.~\cite{arita} 
using {\it ab initio} calculations. In a previous work by some of us~\cite{arita},
indications of hole pairing were shown using 
cluster sizes $L= 2\times 8$. Moreover, the magnetic ordering of the undoped compounds 
involving FM rung and AFM leg was observed, 
as in neutron scattering experiments for BaFe$_2$S$_3$~\cite{nambu}.
However, in those early calculations the DMRG accuracy with regards to number
of states was limited. The present study moves considerably beyond our previous accuracy
by using more modern DMRG codes and computational facilities. Now the cluster size $L=2 \times 12$
is reachable. Even more importantly, considerable progress in the physics is here reported.

In particular, we explore the magnetic properties for various values 
of hole doping and varying interaction strengths over broad ranges. 
We have observed for the first time that the magnetic order could evolve
from the canonical $(\pi,0)$ order to a more exotic $(0,\pi)$ state, with AFM rungs and FM legs.
We have also carefully analyzed the real-space charge density. The number of minima in this quantity
is always half the number of holes, suggesting hole pairing, as along as the magnetic background
remains antiferromagnetic of any kind. The pairs are arranged in what resembles a charge density wave
made of pairs. Remarkably, in regimes where there is no AFM order, such as
in fully ferromagnetic regions at large hole doping and Hubbard strength, pairing disappears.
Moreover, we study pair-pair correlations and find regions where pairing is robust 
(in our previous study~\cite{arita} pair-pair correlations were not addressed).
Here, we have calculated both singlet and triplet pair-correlations using operators defined along the rungs
 of the ladder, and for various hole concentrations. We also study pair correlations employing 
operators along plaquette diagonals, a hole configuration also prominent in the two-hole bound state.
While the rung operator correlations dominate, the diagonal plaquette correlations are similar.

Due to the simultaneous presence of a charge density profile and robust pair-pair singlet correlations,
both in a nontrivial magnetic background, we believe our results can be interpreted in a similar manner
as recent efforts in the context of cuprates within the umbrella of {\it intertwined order parameters}~\cite{inter1,inter2}.
In particular, a recent scenario in the context of cuprates
proposed by Tranquada~\cite{inter3} expresses that latent tendencies toward superconductivity in individual ladders
could lead to an emergent global superconducting state by anti-phase Josephson coupling. While the anti-phase
nature of the coupling is needed in $d$-wave superconductors, in our case simply a mere interladder coupling may be
sufficient for the entire ensemble to become superconducting. Thus, our results raise the exciting
possibility that intertwined order could also be of relevance in iron superconductors, a novel concept.

The organization of the manuscript is as follows. 
Section II contains the two-orbital Hubbard model employed here
for two-leg ladders as well as details of the numerical methods used. 
Section III starts presenting the DMRG results with focus on the various magnetic orders
upon hole doping, as well as the presence of bound states for the case of two holes.
Section IV deals with the pair correlations at various dopings and the real-space
hole distributions, presenting the idea
that intertwined orders could be of relevance for isolated ladders, potentially leading to superconductivity
in an ensemble of weakly coupled ladders.
Finally, in Section V we present our conclusions.

\section{II. Two-Orbital Hubbard Model and Methods} 
\begin{figure}[h]
\centering
\rotatebox{0}{\includegraphics*[width=\linewidth]{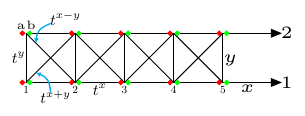}}
	\caption{Schematic representation of a two-leg ladder and the two-orbitals Hubbard model used in this work. At each site, the red circle represents orbital $a$ and the green circle
	orbital $b$. The 2$\times$2 hopping matrix along
	legs is indicated by $t^x$ and along rung by $t^y$. We have also considered
	plaquette diagonal hopping matrices $t^{x+y}$ and $t^{x-y}$. For details see text.
	}
\label{Fig1}
\end{figure}
In this section, we describe the multiorbital Hubbard model 
for the iron-based ladder compound BaFe$_2$S$_3$ used in this work.
The multiorbital Hubbard model for a two-leg ladder can be written 
as the sum of kinetic and interaction energy terms $H=H_k+H_{in}$~\cite{luo}. 
The kinetic portion contains the nearest-neighbor hopping along the leg and rung directions,  
and also next-nearest-neighbor hopping along the plaquette diagonals of the two-leg ladder. Following the convention employed in Ref.~\cite{arita}, the tight binding term is:
\begin{equation}
	H_k = \sum_{i,\sigma,\gamma,\gamma',\vec{\alpha}} t^{\vec{\alpha}}_{\gamma, \gamma'}\left(c^{\dagger}_{i\sigma,\gamma}c^{\phantom\dagger}_{i+{\vec{\alpha}}, \sigma, \gamma'}+H.c.\right) + \sum_{i,\gamma \sigma} \Delta_{\gamma} n_{i,\sigma, \gamma}
\end{equation}

\noindent where $t^{\vec{\alpha}}_{\gamma, \gamma'}$ is the hopping matrix  
along the directions $\vec{\alpha}$ indicated in Fig.~1 and the orbital space includes 
two orbitals $\gamma=\{d_{x^2-y^2},d_{xz}\}$.
For notation simplicity, these two orbitals will be denoted  as $\gamma = \{a,b\}$, respectively. $\Delta_{\gamma}$ denotes the crystal field 
splitting of those two orbitals at $P=12.36$ GPa. 
$n_{i,\sigma, \gamma}$ represents the orbital-resolved number 
operator at site $i$. For our numerical calculations, 
we used the same hopping matrices $t^{\vec{\alpha}}_{\gamma, \gamma'}$
introduced in Ref.~\cite{arita}. 
The actual values of $t^{\vec{\alpha}}_{\gamma, \gamma'}$ were obtained from 
fitting the tight-binding bands with the {\it{ab initio}} downfolded band structure calculations~\cite{arita}.
The 2$\times$2 hopping matrix between sites $i$ and $i+{\hat x}$ 
along the legs of the ladder $t^{x}_{\gamma, \gamma'}$ is given by (in eV units):
\[
t^{x}_{\gamma, \gamma'}=
  \begin{bmatrix}
    -0.334 & -0.177  \\
    +0.177 & +0.212  
  \end{bmatrix}
\] 
where $\gamma$ is the orbital index for site $i$ and  $\gamma'$ for $i+{\hat x}$ . $t^{y}_{\gamma, \gamma'}$ is the 2$\times$2 hopping matrix along the vertical rung-direction:
\[
t^{y}_{\gamma, \gamma'}=
  \begin{bmatrix}
    -0.024 &  0.000 \\
     0.000 & +0.216 
  \end{bmatrix}
\]
The $t^{x+y}$ and $t^{x-y}$ are 2$\times$2 hopping matrices along the plaquette diagonals of the ladder:
\[
t^{x+y}_{\gamma, \gamma'}= t^{x-y}_{\gamma, \gamma'}=
  \begin{bmatrix}
    +0.085 & +0.216 \\
    -0.216 & +0.109 
  \end{bmatrix}
\]
The crystal fields $\Delta_{\gamma}$  at $P=12.36$~GPa for each orbital are $\Delta_{a}=0.423$~eV 
and $\Delta_{b}=-0.314$~eV. The kinetic energy bandwidth is $W=2.2533$~eV.

The electronic interaction portion of the Hamiltonian is:
\begin{eqnarray}
H_{in}= U\sum_{i\gamma}n_{i\uparrow \gamma} n_{i\downarrow \gamma} +\left(U'-\frac{J_H}{2}\right) \sum_{i,\gamma < \gamma'} n_{i \gamma} n_{i\gamma'} \nonumber \\
-2J_H  \sum_{i,\gamma < \gamma'} {{\bf S}_{i,\gamma}}\cdot{{\bf S}_{i,\gamma'}}+J_H  \sum_{i,\gamma < \gamma'} \left(P^+_{i\gamma} P_{i\gamma'}+H.c.\right). 
\end{eqnarray}
The first term is the standard on-site Hubbard repulsion between $\uparrow$ and $\downarrow$ electrons in the same orbital. 
The second term contain the on-site electronic repulsion between electrons at 
different orbitals and same site. As often employed in previous publications,
the standard relation $U'=U-2J_H$ is here assumed, due to the $SU(2)$ symmetry of the Hamiltonian. Also
the widely employed ratio $J_H/U=0.25$ is here also used because iron superconductors are known to have
a relatively large Hund coupling. The third term is the ferromagnetic 
Hund's interaction between electrons occupying the active 
two orbitals $\gamma=\{d_{x^2-y^2},d_{xz}\}$. 
The operator ${\bf S}_{i,\gamma}$ is the total spin of orbital $\gamma$ 
at site $i$. The last term is the 
pair-hopping between different orbitals at site $i$,
where $P_{i \gamma}$=$c_{i \downarrow \gamma} c_{i \uparrow \gamma}$. This pair-hopping term arises
from Coulomb interaction matrix elements and has no influence on the more extended pairing of holes
due to magnetic short-range order discussed in this publication. Moreover,  to confirm this perspective,
in many examples we turned off the on-site pair hopping term in the Hamiltonian and our  results were barely modified. 

To solve this two-orbital Hubbard model for the ladder compound BaFe$_2$S$_3$  we employed the  DMRG method. 
Here we focused on a cluster of size $L= 2 \times 12$
and with various values for the hole doping. The number of holes $N_h$ 
was obtained by removing $N_h$ electrons from the half-filled
system with $N=48$ electrons (i.e. 24 sites and 2 electrons per site due to the two orbitals).
Due to recent improvement in our computational capabilities, 
we are able to keep up to $m=3800$ states for our DMRG calculation. 
As a result we can perform the DMRG calculation quite accurately, 
with various hole doping concentrations employed for the $L= 2 \times 12$ cluster. 
To characterize magnetic and superconducting properties, 
we have calculated various observables, such as the
charge and spin correlations, spin structure factors, and the pair-pair correlations.
For the DMRG calculations we employed open-boundary conditions and we 
used the DMRG++ software~\cite{gonzalo}. 

\section{III. Magnetic order in doped ladders}
\subsection{A. Effect of doping on magnetic ordering at intermediate Hubbard coupling}
\begin{figure}[h]
\centering
\rotatebox{0}{\includegraphics*[width=\linewidth]{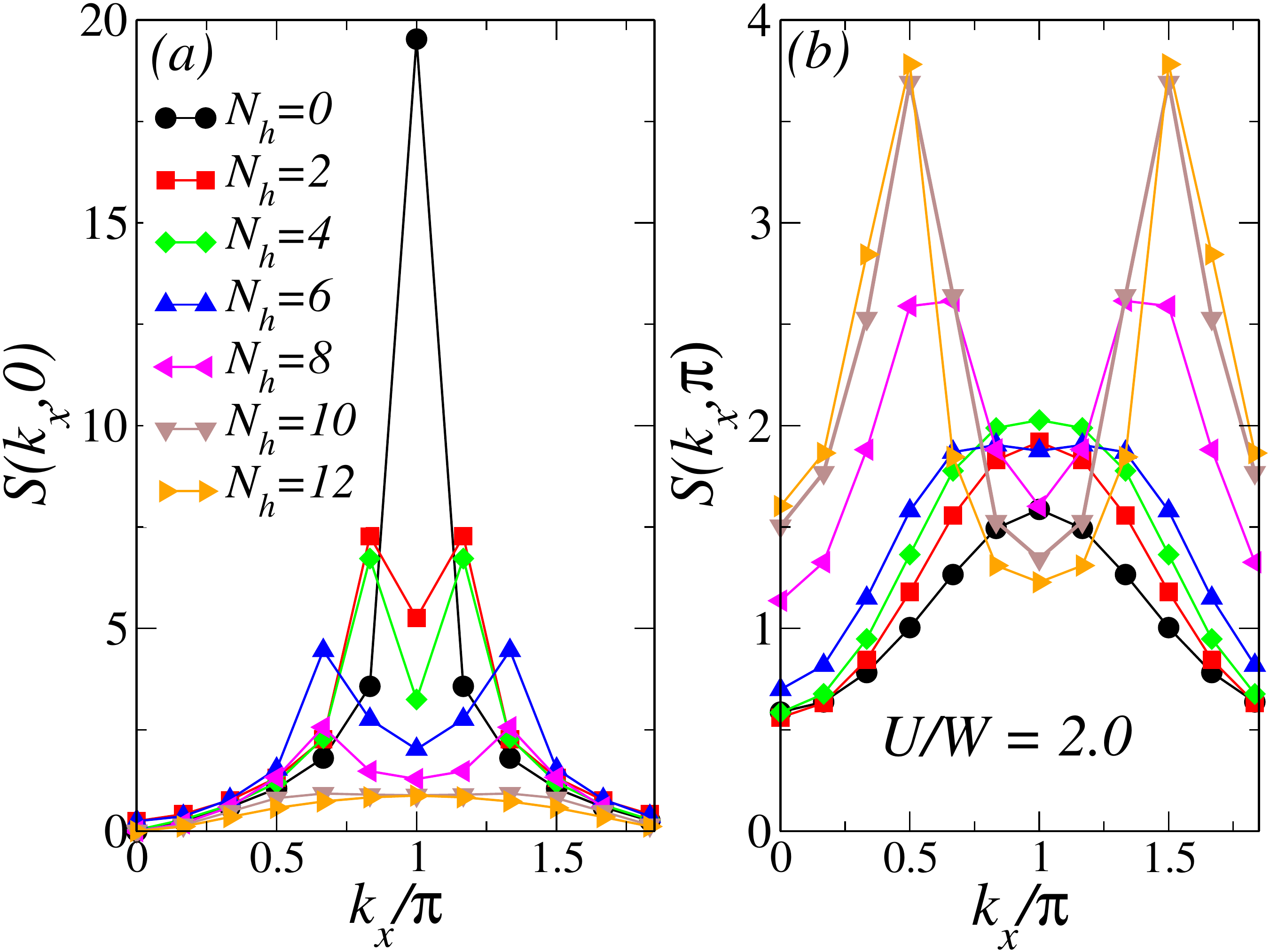}}
\rotatebox{0}{\includegraphics*[width=\linewidth]{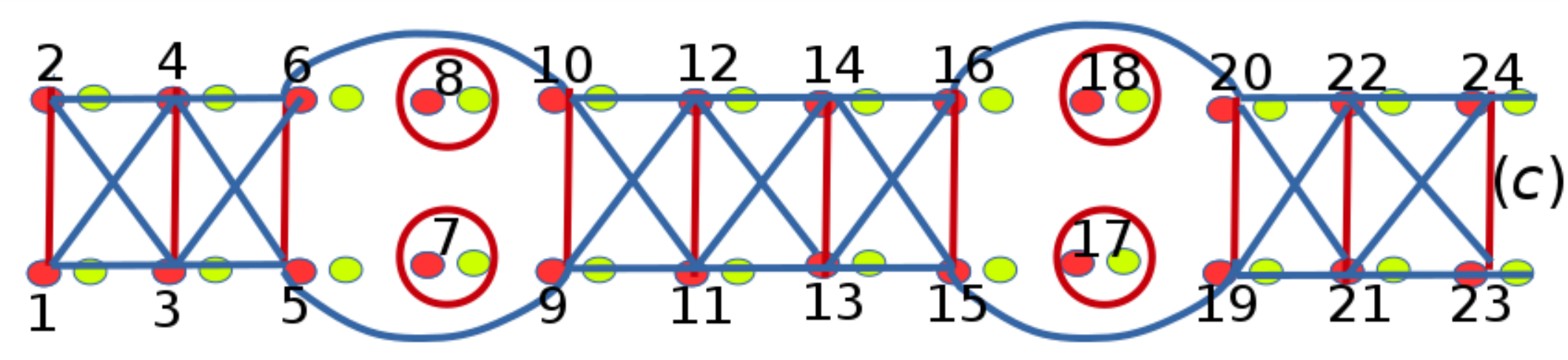}}
\rotatebox{0}{\includegraphics*[width=\linewidth]{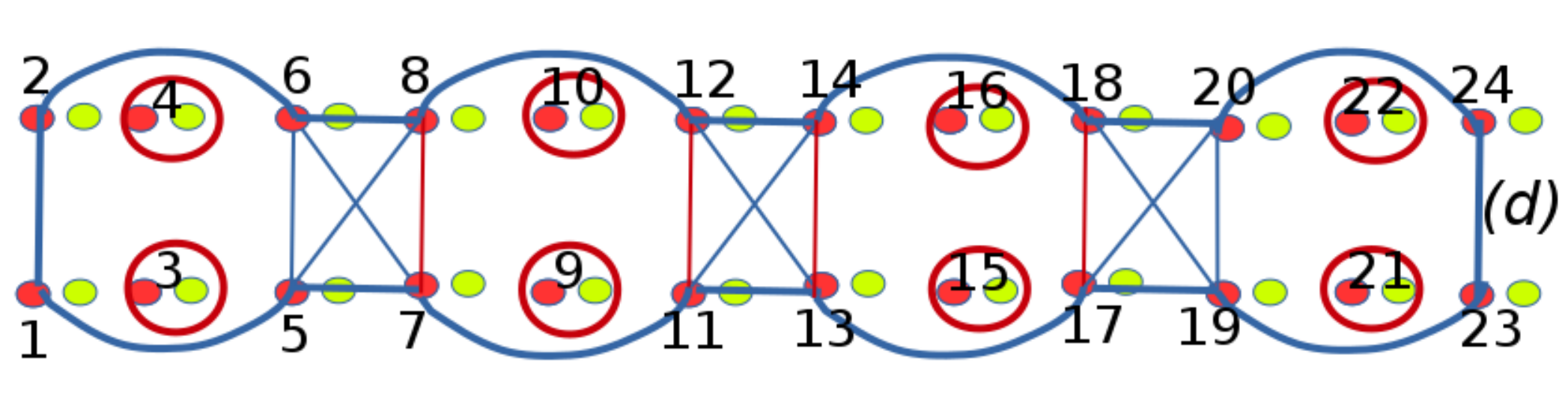}}
	\caption{Spin structure factors (a) S($k_x$,0) (b) S($k_x$,$\pi$)
	vs wave vector $k_x$ for various values of hole doping.
	(c,d) Spin-spin correlation at a  fixed and site-projected arrangement of 
	holes along the rungs of the ladder.
	(c) is for four holes projected at the four red circles indicated. 
	(d) is for eight holes projected at the eight red circles.
	Blue color lines represent AFM bonds, whereas FM bonds are
	represented with red color. These results were obtained 
	using DMRG at $U/W=2.0$ and $J_H/U=0.25$, for cluster 
	size $L=2 \times 12$. 
	}
\label{Fig2}
\end{figure}

For iron-based high-$T_c$ compounds, it is widely believed that 
magnetic-fluctuations play a crucial role 
to induce superconductivity~\cite{dai,dagotto}.
For this reason, it is important to determine  
the magnetic ordering varying the hole doping, starting with the undoped state. 
At half-filling BaFe$_2$S$_3$ is a Mott 
insulator and displays stripe-type magnetic ordering, namely antiferromagnetic order along the
legs and ferromagnetic along the rungs~\cite{nambu}. 
When changing the coupling values and hole concentrations~\cite{xavier}, 
there are possibilities of various types of magnetic ordering,
 due to the competition between the kinetic energy of holes,
the superexchange between local Fe moments, and the Hund coupling.
To find the magnetic ordering for different number of holes 
and interaction strength $U/W$, 
we calculate the spin-structure factor $S(k_x,k_y)=\frac{1}{L} \sum_{i,j} e^{{\bf{k}} \cdot {\bf{r}}_{ij}} \langle {\bf{S}}_i \cdot {\bf{S}}_j \rangle$ 
for cluster size $L= 2 \times 12$ for the allowed set of wave vectors {$(k_x,0)$ and $(k_x,\pi)$}.

Figure~\ref{Fig2}(a) contains the spin structure factor $S(k_x,0)$ 
vs. wave vector $k_x$ at $U/W=2.0$ for various values of hole doping. 
At half filling ($N_e =48$ in the $L=2\times 12$ cluster), 
$S(k_x,0)$ displays a sharp peak at $(\pi,0)$, 
which is equivalent to AFM-order in the $x$-direction (leg) and FM-order along the 
$y$-direction (rung) of the ladder. This type of magnetic ordering is compatible with
neutron experiments for BaFe$_2$S$_2$ at ambient pressure~\cite{nambu}.
Interestingly, increasing the number of holes, 
the peak of $S(q_x,0)$ start splitting and the height of the peak decreases.
This peak splitting at ($\pi$,0)
increasing with the hole concentration indicates the appearance of
spin incommensurability (IC) in the system~\cite{patel1}. 
This spin incommensurate spin fluctuations
have been observed in the hole-overdoped iron-based high-$T_c$ compound
KFe$_2$As$_2$ ~\cite{lee}. The decrease in intensity of  the peak at 
 ($\pi$,0) is due to the scrambling of spin-ordering by the holes, and also by the reduction
in the number of electrons. 
Eventually, for large enough hole doping, $S(k_x,0)$ displays no prominent ordering,  
while $S(k_x,\pi)$ shows a commensurate peak at ($\pi/2,\pi $), as  shown in Fig.~\ref{Fig2}(b).  
At half-filling and for $U/W \gtrsim 1$ and $J_H/U=0.25$,
the local spin moments are fully developed
and approach their maximum value $S=1$ at each site for the two-orbital model used here
(corresponding to magnetic moment $2.0 \mu_B $)~\cite{arita}.
We find that the averaged spin-square expectation value 
$\langle S^2 \rangle = \frac{1}{L}\sum_i \langle 
{\bf{S}}_i \cdot {\bf{S}}_i\rangle $ decreases 
linearly with  hole doping ($N_h$) [see inset of Fig.~\ref{Fig3}(b)], 
again suggesting the scrambling of local spin order with increasing holes $N_h$.

To better visualize in real space the spin IC and scrambling of spin-ordering 
when increasing the  number of holes $N_h$, we have calculated the spin-spin 
correlations when the holes are projected into their most probable
locations~\cite{gazza,martin}. 
We found that the
holes primarily reside in orbital $a$ due to the crystal-field splitting. For each number of holes,
we focused on the configuration of holes with the highest probability in the wave function.
In this case, to project out the dominant configuration of $N_h$ holes residing
on orbital $a$, we use the projection operator~\cite{scalapino} $P_{h a}(h)= P_{ha}(h_1) P_{ha}(h_2) ...P_{ha}(h_m)$ [with $P_{ha}(i)= c_{ia \uparrow} c_{ia \uparrow}^{\dagger} c_{ia \downarrow} c_{ia \downarrow}^{\dagger}$ 
acting as projectors on  the ground state, namely finding the portion of the wave function 
where site $i$ and orbital $a$ are vacant (hole)].

After projecting the holes into their most probable locations,
we calculate the local spin-spin correlations~\cite{gazza,martin,arita}
$\langle \Psi| {\bf{S_{i}}}a \cdot {\bf{S_{j}}}a P_{ha} \Psi\rangle/ \langle \Psi| P_{ha} \Psi\rangle$. In Ref.~\cite{arita} the authors explored 
the spin-spin correlations for 1 and 2 holes at $U/W=2.0$, 
where they found robust AFM-correlations ``across the holes''. 
Here in Fig.~\ref{Fig2}(c) and Fig.~\ref{Fig2}(d), 
we analyze the spin-spin correlation for the cases of 4 and 8 holes, respectively, which are of more
relevance for our focus on pairing. 
We find at $U/W=2$ that with 4  and 8 holes
the spin AFM-correlation ``across the holes'' is still robust. 
The AFM-correlations in the leg direction and FM-correlations along-the rung, characteristic of the undoped
regime, are mostly preserved in locations away from the hole positions. 
However, near the holes,
 AFM-correlations appear across the holes, 
 leading to spin-IC tendencies and broadening of
 the ($\pi,0$) peak.

\subsection{B. Evolution of magnetic order with increasing $U/W$ towards the
strong coupling regime}
\begin{figure}[h]
\centering
\rotatebox{0}{\includegraphics*[width=\linewidth]{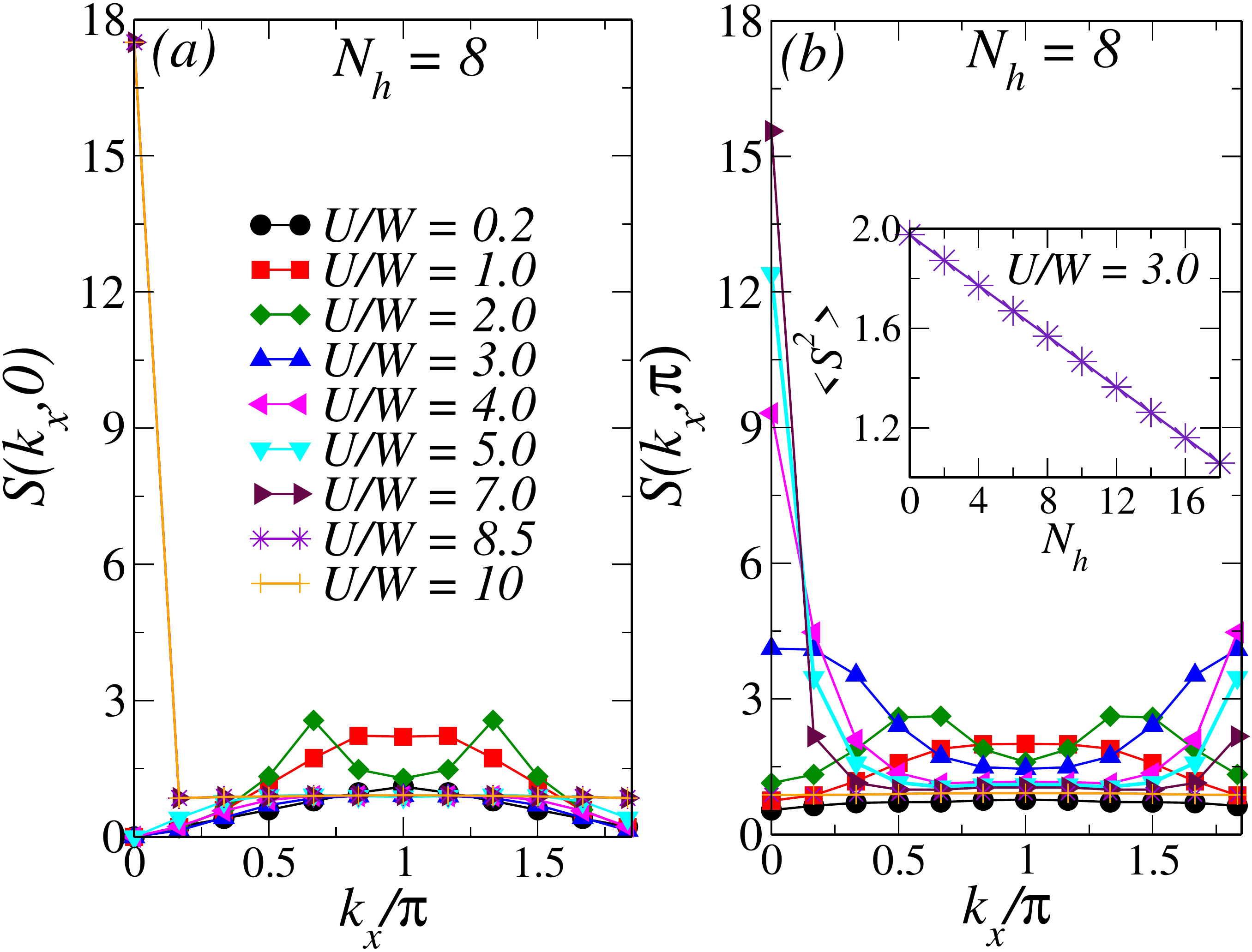}}
\rotatebox{0}{\includegraphics*[width=\linewidth]{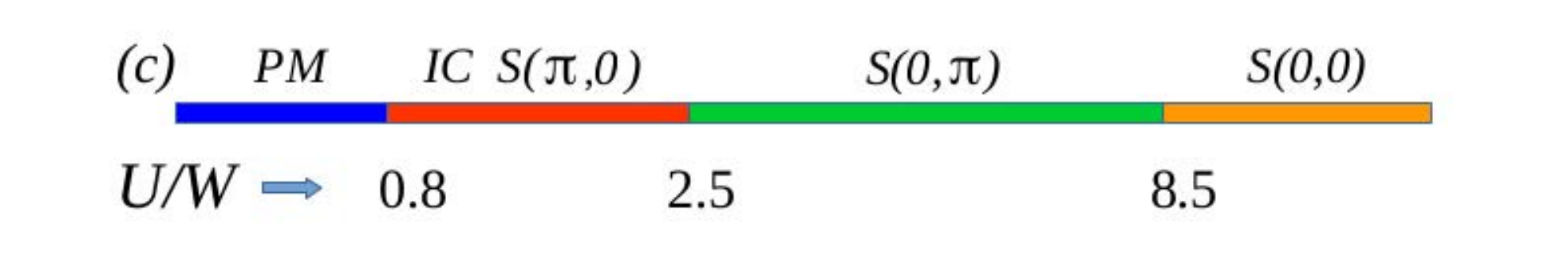}}
	\caption{Spin structure factor (a) S($k_x$,0) (b) S($k_x$,$\pi$)
        vs wave vector $k_x$ for various values of interaction strength $U/W$
	 at fixed number of holes $N_h=8$. Inset shows site averaged 
	 $\langle S^2 \rangle$ vs number of holes $N_h$ at $U/W=3.0$. 
	 (c) Sketch of the magnetic phase diagram at fixed $J_H/U=0.25$ and $N_h=8$.
	}
\label{Fig3}
\end{figure}
 Figure~\ref{Fig3} displays the spin structure factors 
 $S(k_x,0)$ and $S(k_x,\pi)$
 for different values of $U/W$ and at fixed number of holes $N_h=8$. 
 In the weak-coupling limit $0.2 \lesssim U/W \lesssim 0.8$,
  the spin structure factor $S(k_x,0)$ shows no prominent spin-ordering.
  In the range $0.8 \lesssim U/W \lesssim 2.5$,
  $S(k_x,0)$ shows incommensurate spin ordering, where we
  find short-range AFM-spin correlation along legs and FM-correlation 
  along rungs of the ladder. Further increasing $U/W$, 
  we find that the magnetic ordering evolves continuously and eventually
  ferromagnetic tendencies emerge along the leg direction.
  For interaction strength
  $2.5 \lesssim U/W \lesssim 8.5$ the $S(0,\pi)$ ordering, opposite to the previously discussed $S(\pi,0)$ at smaller $U/W$, dominates 
  [see Fig.~\ref{Fig3}(b)], 
  which is equivalent to FM-spin ordering in the leg direction and 
  AFM-spin ordering along the rung of the ladder. 
\begin{figure}[h]
\centering
\rotatebox{0}{\includegraphics*[width=\linewidth]{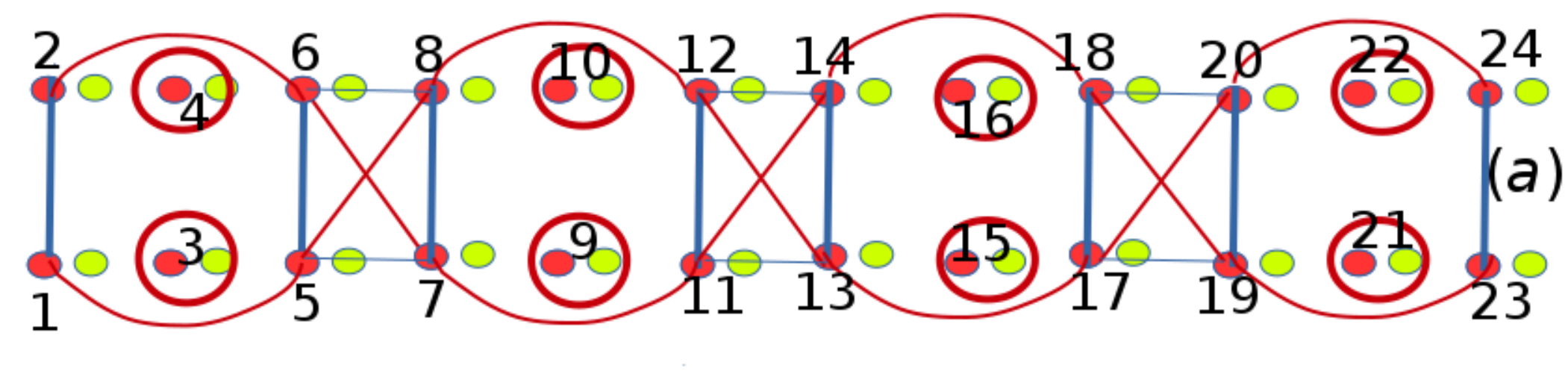}}
\rotatebox{0}{\includegraphics*[width=\linewidth]{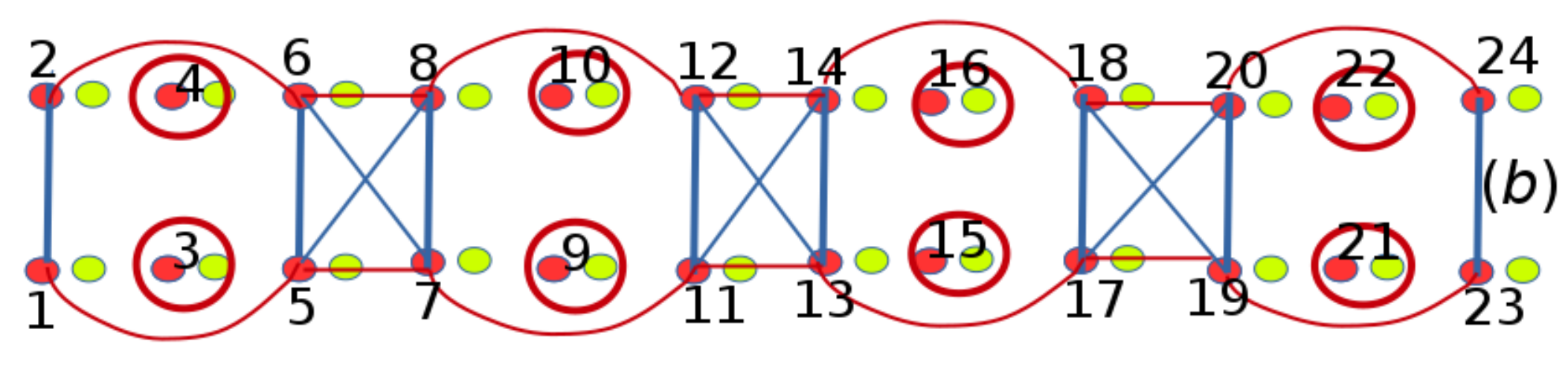}}
\rotatebox{0}{\includegraphics*[width=\linewidth]{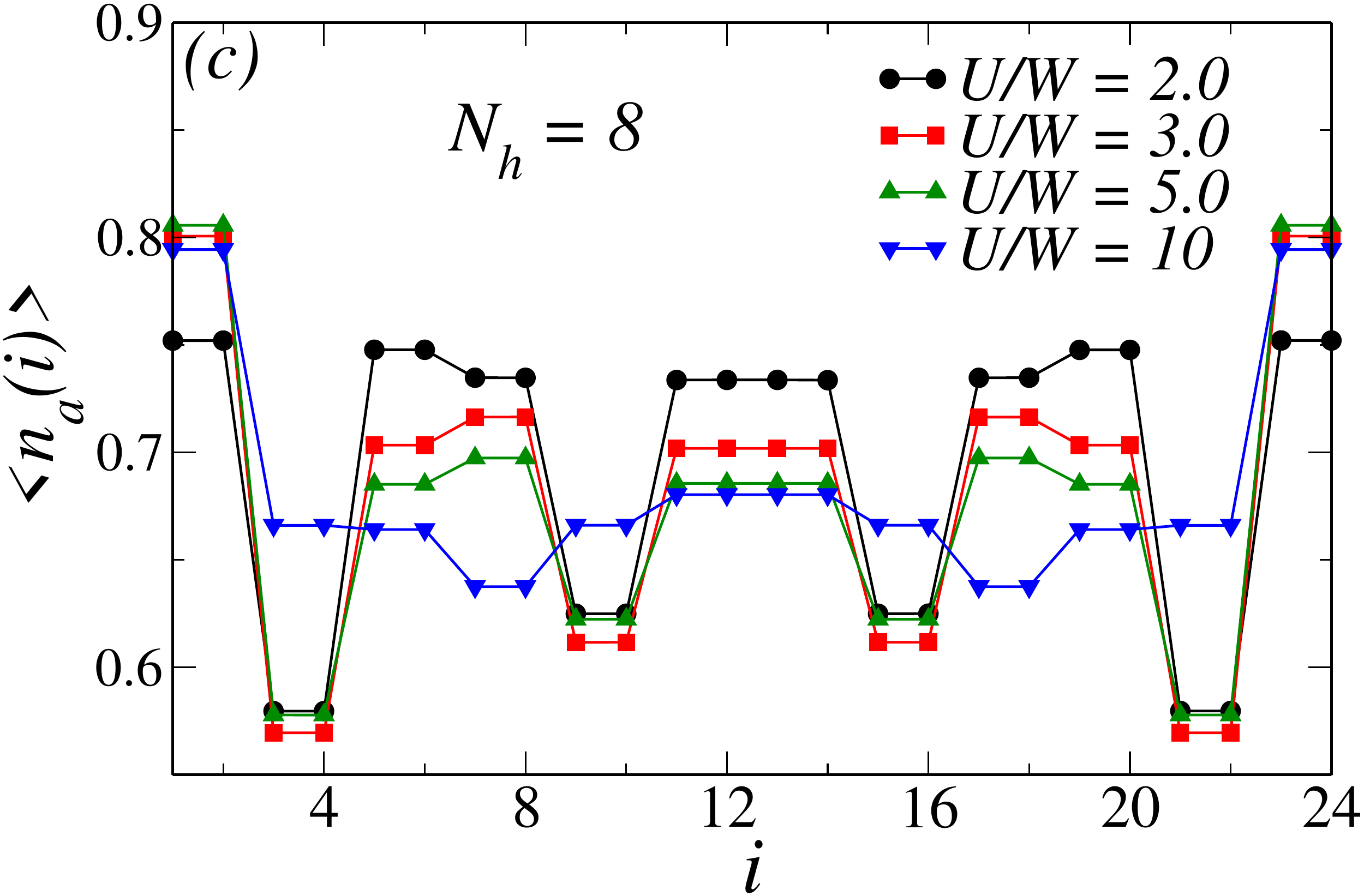}}
	\caption{(a,b) Spin-spin correlation at a fixed and site-projected 
	arrangement of $N_h=8$ holes along the rungs of the ladder.
	  (a) at $U/W=3$, (b) at $U/W=5$.
	 Blue color lines represent AFM bonds, whereas FM bonds are
        represented with red color.
	  (c) Real-space density $\langle n_a(i) \rangle$ vs site index $i$ 
	  with $N_h=8$ holes and for various values of $U/W$.
	}
\label{Fig4}
\end{figure}

\begin{figure}[h]
\centering
\rotatebox{0}{\includegraphics*[width=\linewidth]{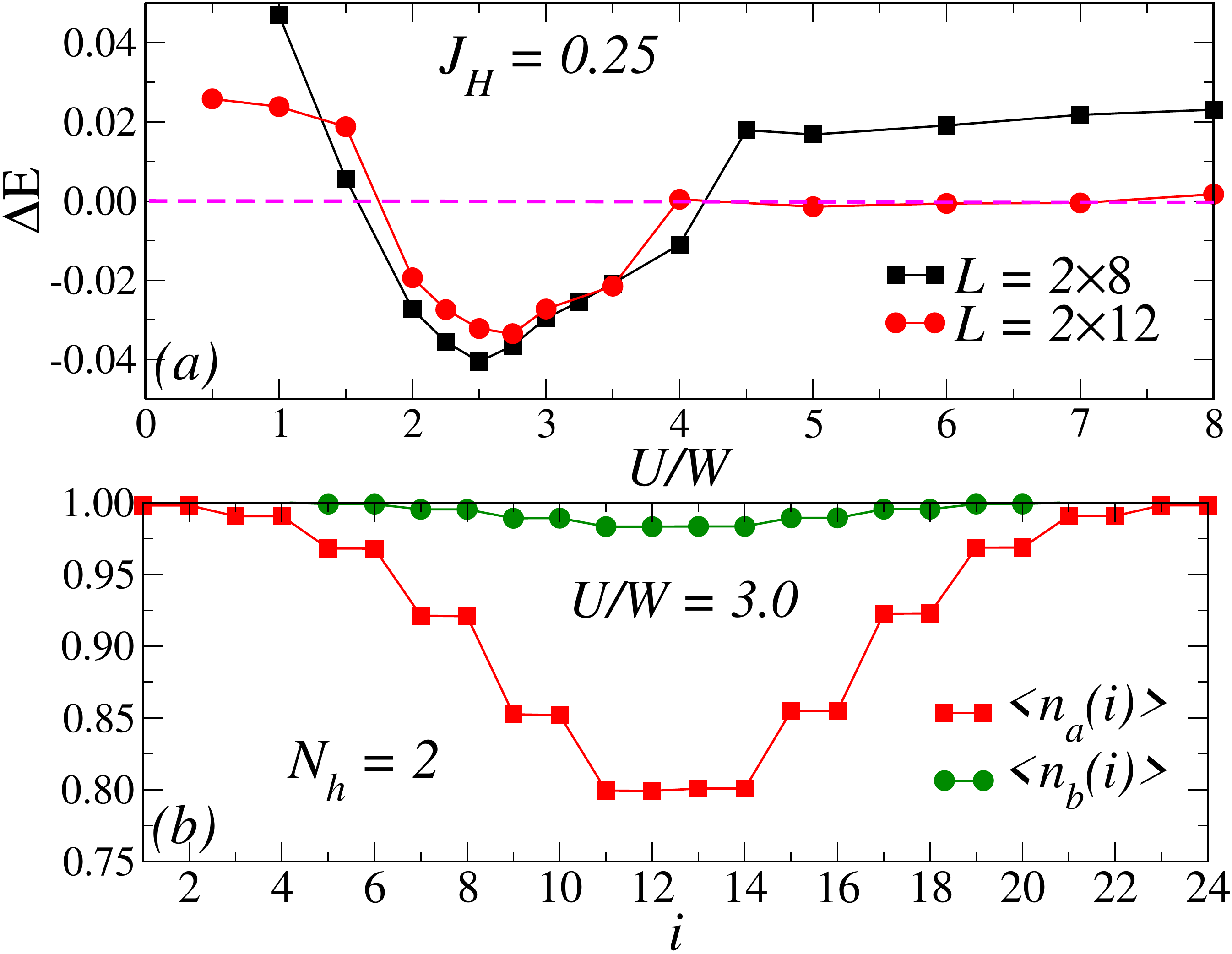}}
\rotatebox{0}{\includegraphics*[width=\linewidth]{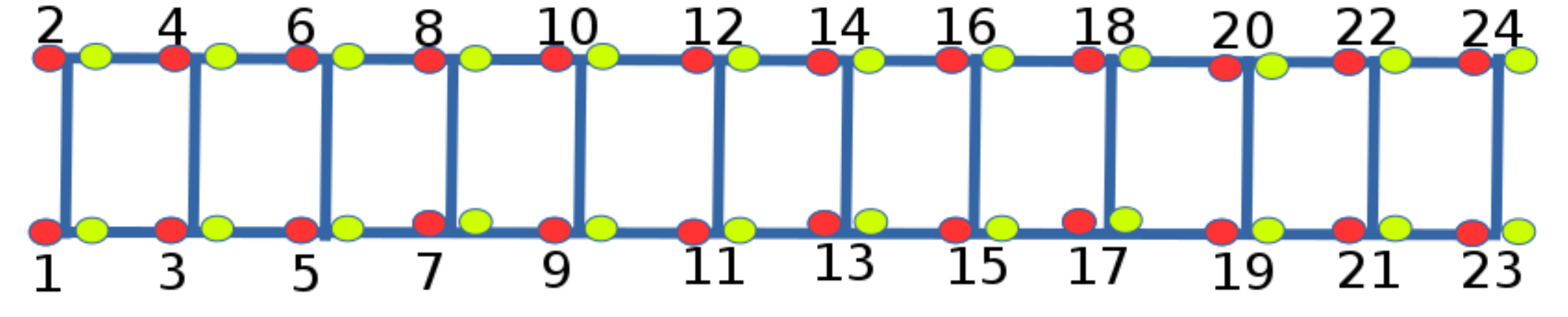}}
	\caption{(a) Binding energy $\Delta E$ vs. interaction strength $U/W$
	at $J_H/U=0.25$ for cluster sizes $L=2 \times 8$ and $2 \times 12$ for comparison. 
	(b) Real-space charge density $\langle n_{\alpha} \rangle$ vs site index $i$ for the case of 
$N_h=2$ holes and at $U/W=3.0$. 
 Schematic representation of a two-leg ladder with the snake-like counting of 
	sites appropriate for DMRG. 
	}
\label{Fig5}
\end{figure}

Surprisingly, for $U/W\gtrsim 8.5$ the structure factor $S(k_x,0)$
 shows a strong peak at wave vector (0,0) [Fig.~\ref{Fig3}(a)], 
 indicating the appearance of  ferromagnetic spin-order both along the leg and 
 the rung directions of the ladder.
 As already explained, holes mainly reside on orbital $a$, 
 while the population of orbital $b$ stays close to one electron/site 
 (see Fig.~\ref{Fig5}). For this reason the orbital $b$ behaves as a localized 
 spin 1/2 system, while orbital $a$ provides delocalized holes. 
 The movement (kinetic energy) of holes can improve if all spins aligned 
 in the same direction~\cite{xavier}, particularly at large $U/W$ because the Hund
coupling is growing linearly with the Hubbard strength via $J_H/U=0.25$.
 This scenario is compatible with the double-exchange mechanism for manganites and  
 results in ferromagnetic tendencies with increasing $U/W$.
 We find that for large doping the ferromagnetic tendency appears 
 at smaller values of $U/W$ in comparison to the low-hole doping case~\cite{arita}.

The above described numerical results show that the magnetic order changes
with increasing $U/W$.
To understand the connection between magnetic ordering and pairing of holes, 
in Fig.~\ref{Fig4} we analyze the projected spin correlation $\langle \Psi| {\bf{S_{i}}}a \cdot {\bf{S_{j}}}a P_{ha} \Psi\rangle/ \langle \Psi P_{ha} \Psi\rangle$ and the real-space charge
density at orbital $a$ for different values of $U/W$.
Figure~\ref{Fig4}(a) shows the projected spin correlations for $N_h=8$
at $U/W=3.0$. The FM magnetic order along the rung found at $U/W=2.0$ [Fig.~\ref{Fig2}(d)]
now has changed to AFM order along the same rungs at $U/W=3.0$.
Along the leg direction, at $U/W=2.0$ the order was fully AFM but at $U/W=3.0$ now
there is a mixture: some bonds are FM and others are AFM. Once the FM vs AFM intensities are
added, overall the leg magnetic ordering becomes ferromagnetic,
consistent with the previously shown spin structure factor $S(0,\pi)$ [Fig.~\ref{Fig3}(b)].
Interestingly, at $U/W=5.0$ [Fig.~\ref{Fig4}(b)] the magnetic
order along the legs switches completely to a FM order, 
whereas along the rung it is completely AFM order, i.e. 
the ordering is fully reversed as compared to $U/W=2.0$
where the rungs are FM and legs are AFM.
For this reason at this coupling we find a peak at 
$(0,\pi)$ for the spin structure factor $S(k_x,\pi)$. 

Figure~\ref{Fig4}(c) shows the real-space charge density of orbital $a$ 
for different values of $U/W$ with $N_h=8$ holes. 
It is clear from Fig.~\ref{Fig4}(c) that
pairs still exist, and they are still primarily located 
along the rungs of the ladder, at all values of $U/W$ where there is
a mixture of FM and AFM tendencies, namely whether favoring $(0,\pi)$ or $(\pi,0)$ ordering.
On the other hand, both in the limits of small $U/W$ with no magnetic order [Fig.~\ref{Fig6}(c)] 
and at very large $U/W$ such as
$U/W=10.0$ [see Fig.~\ref{Fig4}(c), where the order is FM in both the rung and leg directions] we find that the 
density remains approximately uniform and there are no indications of pair formation because the density
profile does not show four minima for the case of $N_h=8$.

\section{IV. Pairing Tendencies of Holes}

\subsection{A. Formation of holes pairs}

In this subsection, we investigate the pairing tendencies after the addition of
holes to the undoped state. The formation of hole-pairs 
is a precursor of a pair-density-wave or of superconductivity~\cite{xavier,patel1,arita}. Recent
investigations indicate that in the high-$T_c$ cuprates, the region of pair-density-wave could become globally
superconducting by a weak coupling between them~\cite{inter3}.
 
In order to find the  hole-pairing region as a function of 
interaction parameter $U/W$, we calculated the binding energy $\Delta E $. 
The binding energy of a pair of holes is defined as 
$\Delta E = E(N-2) +E(N)-2E(N-1)$,
where $E(N)$ is the ground state with $N$ electrons~\cite{patel1,arita}. 
For a finite cluster,  $\Delta < 0$ is indicative of bound state of holes, 
while $\Delta > 0$, or approximately zero, suggests two holes do not form a bound state.
In Fig.~\ref{Fig5}(a), we show the binding energy for two cluster 
sizes $L= 2 \times 8$ and $L= 2 \times 12$ vs. interaction 
parameter $U/W$ and at fixed $J_H/U=0.25$.
Interestingly, the binding energies $\Delta$ 
become negative in the region $ 1.6 \lesssim U/W \lesssim 4.0$, 
indicate of formation of stable bound pairs of holes in this regime.
The similarity among the values of binding energies for both cluster sizes $L= 2 \times 8$,
and $ 2\times 12$ indicate small finite size effects in the bnding region. Note that
in the bulk limit, when there is no pairing the binding energy should be zero because the
energy of 2 holes is the same as 2 times the energy of 1 hole, measured with respect to the undoped
ground state. The case $2 \times 12$ already shows this behavior in a broad range of robust $U/W$.
At weak coupling, where size effects are often larger than in strong coupling, the tendency increasing $L$ 
is in the right direction: $2\times 12$ has smaller binding energy than $2 \times 8$ before true binding occurs.

 To further verify the pairing of holes,  
we analyze the real-space charge density $\langle n_{\alpha} \rangle$ 
 with two holes at $U/W=3.0$. As shown in Fig.~\ref{Fig5}(b), 
 the occupancy of orbital $b$ is close to one, 
 while there is a deep minimum in the density of orbital $a$. This indicates that 
 holes reside mainly at orbital $a$. This charge profile is also compatible with a bound state,
as opposed to two unbounded holes where two minima would be expected.
\begin{figure}[h]
\centering
\rotatebox{0}{\includegraphics*[width=\linewidth]{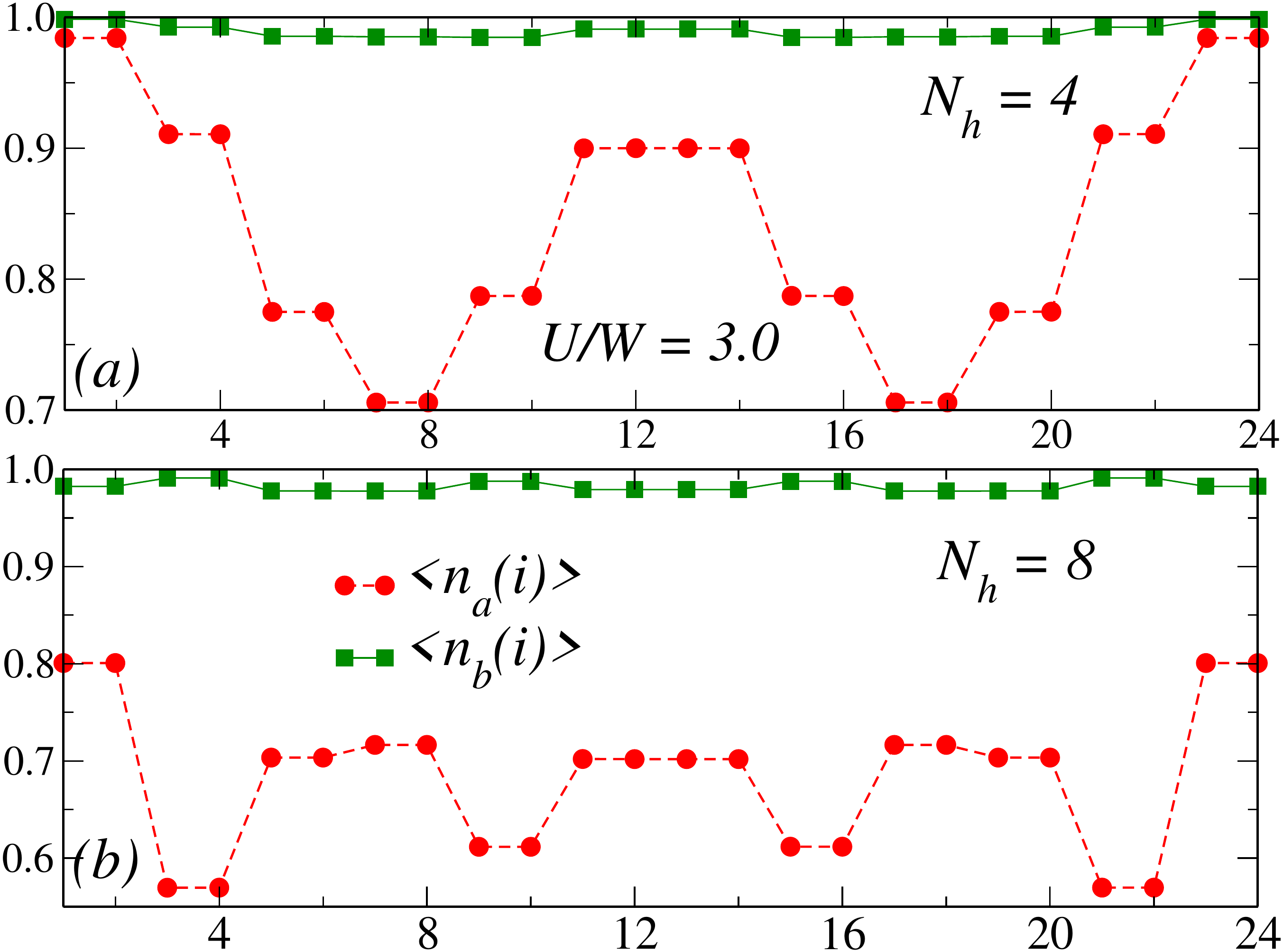}}
\rotatebox{0}{\includegraphics*[width=\linewidth]{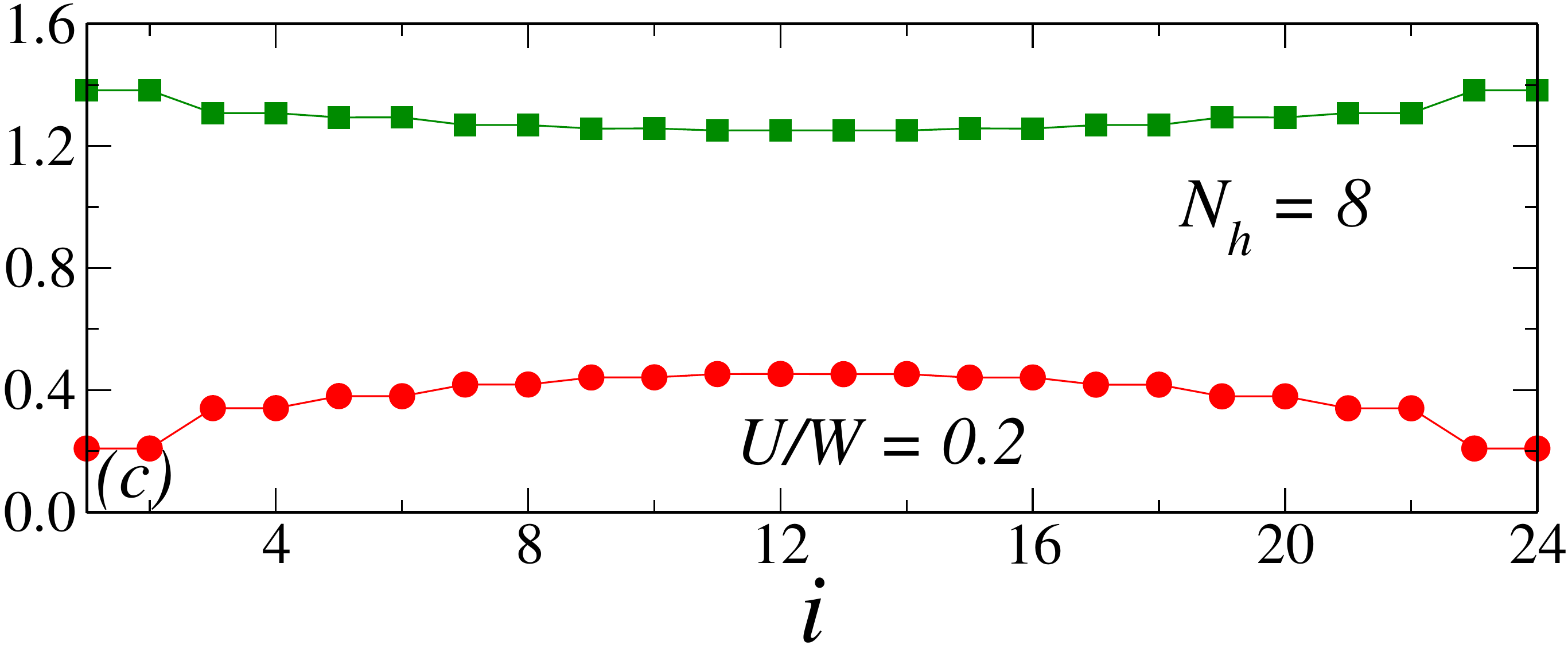}}
	\caption{Real-space density profile at $U/W=3.0$ (a)
         $N_h=4$ holes and for (b) $N_h=8$.holes.
	(c) Real-space density profile at $U/W=0.2$ with $N_h=8$ holes. 
	These numerical calculations were
	performed using DMRG for a cluster
	size $L=2 \times 12$ and at fixed $J_H/U =0.25$. 
	}
\label{Fig6}
\end{figure}
 The real-space density $\langle n_a(i) \rangle$ of orbital $a$, 
 also suggest that binding of holes occurs near the center of the cluster,
 either along the same rung or along the plaquette diagonals,
 consistent with the negative value of binding energy at $U/W=3.0$.

To understand the pairing tendencies for more than two holes,
in Figs.~\ref{Fig6}(a) and (b),
we plot the real-space charge densities for each orbital
with $N_h= 4$ and $N_h=8$ holes  at $U/W=3.0$. 
The densities of orbital $b$ remain approximately one,
and holes reside mainly on orbital $a$. 
These  tendencies resemble the orbital selective Mott phase discussed extensively 
in recent literature related to ladders and chains~\cite{rincon,OSMP0,OSMP1,OSMP2,OSMP3}, 
at doping $N_h=4$ and $N_h=8$.

For orbital $a$ and with 4 holes Fig.~\ref{Fig6}(a), there exist two minima
in the real-space charge density, compatible with two bound pairs of holes. 
These four holes are primarily located along the rungs 
 with site indexes [$i=7$,$8$] and [$i=17$,$18$] in the snake-geometry convention.
For the case of $N_h=8$ holes there exist four minima [see Fig.~\ref{Fig6}(b)],
and these minima are also located with the highest chances along the rungs of the ladder.
Figure~\ref{Fig6}(c) shows the density of orbital $a$ and $b$ 
at lower values of the interaction $U/W=0.2$, for the case of $N_h=8$ holes.
In this weak coupling example, the density of orbitals, both $\langle n_a \rangle $ and $\langle n_b \rangle $,
takes nearly uniform values. There are no minima in the density profile,
compatible with no bound pairs formation in this range of $U/W$.
The results above suggest that for larger hole doping,  
holes form bound pairs along 
the rungs of the ladder and these holes reside mainly on orbital $a$, result compatible
with our previous investigations.

\subsection{B. Superconducting Correlations}
\begin{figure}[h]
\centering
\rotatebox{0}{\includegraphics*[width=\linewidth]{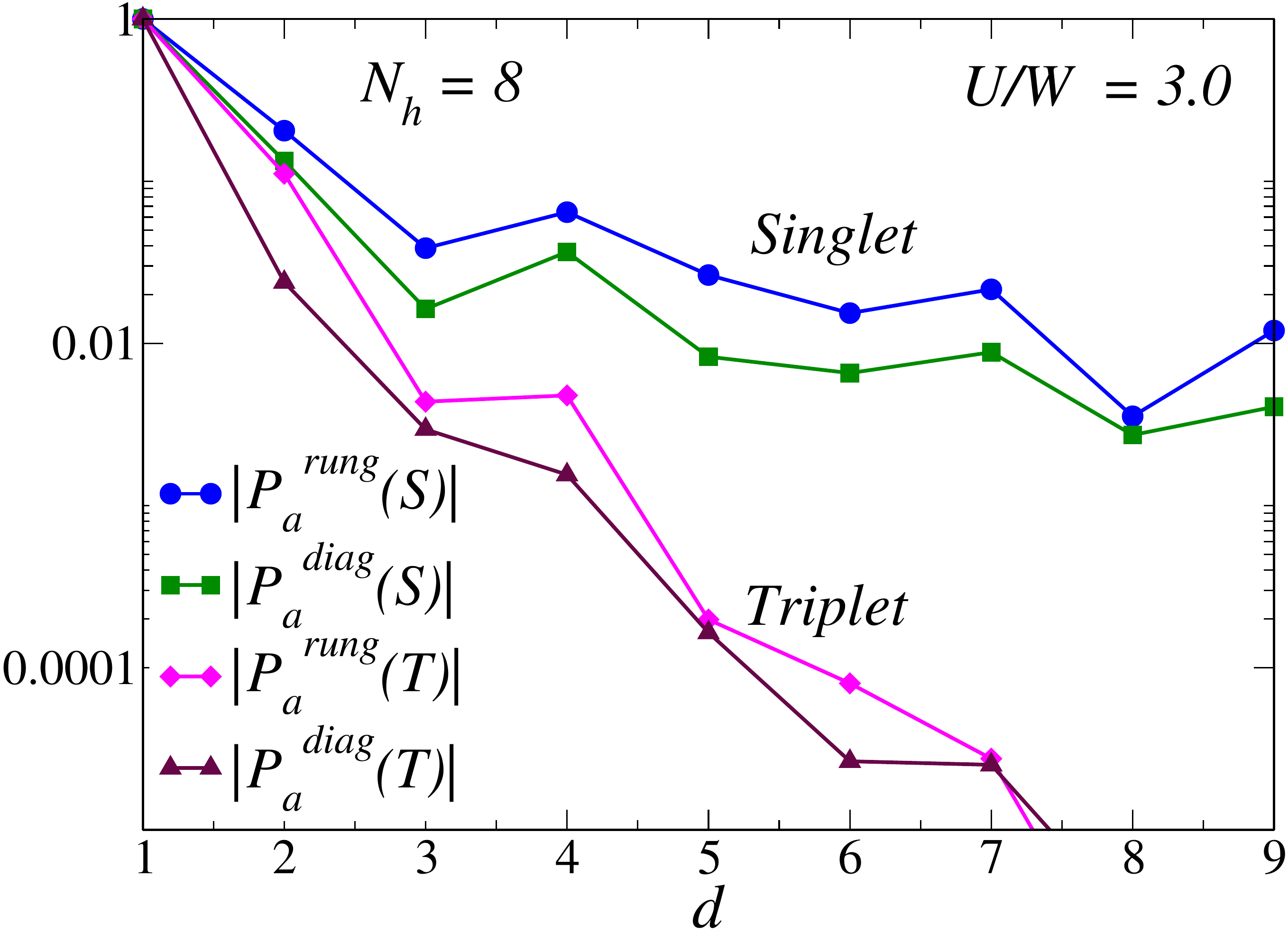}}
\rotatebox{0}{\includegraphics*[width=\linewidth]{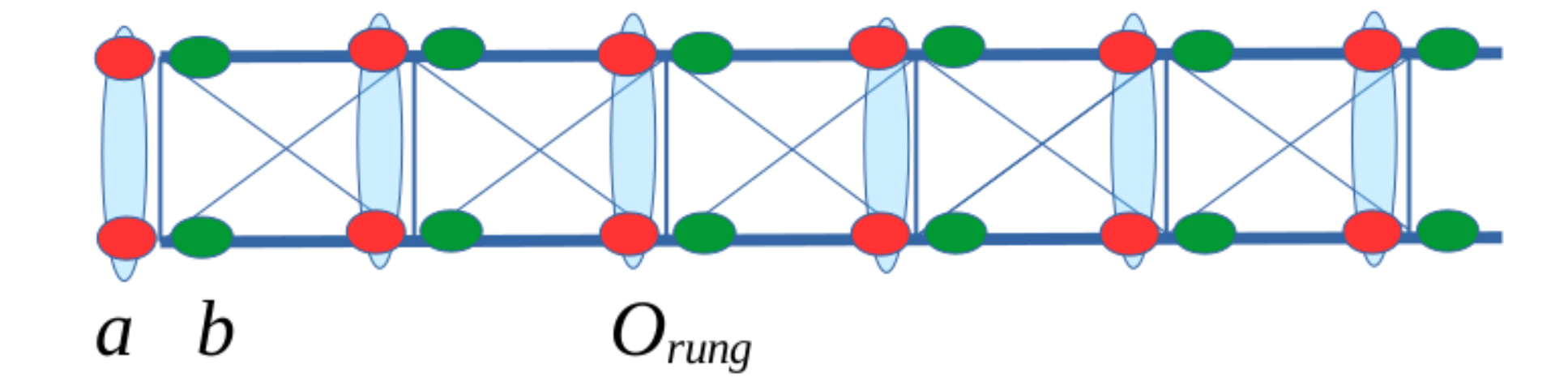}}
\rotatebox{0}{\includegraphics*[width=\linewidth]{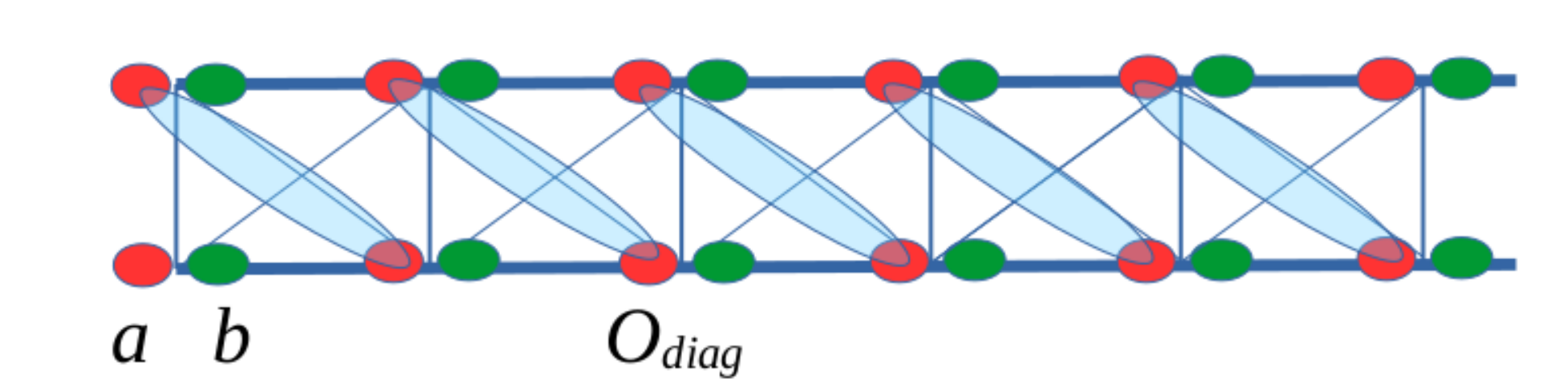}}
	\caption{Singlet and triplet pair-correlations, 
	$P_a^{rung}(S)$ and $P_a^{rung}(T)$, calculated along the 
	vertical rung, whereas $P_a^{diag}(S)$ and $P_a^{diag}(T)$ 
	are calculated along diagonal of the ladder, as indicated in the sketch at the bottom of the figure.
All results obtained at $U/W=3.0$ and $J_H/U=0.25$ for a size cluster $L=2 \times 12$.
	Bottom: schematic diagrams of vertical $O_{rung}$ and diagonal 
	$O_{diag}$ pairing operators involving orbital $a$.
	}
\label{Fig7}
\end{figure}

Superconductivity involves pairing of holes, as in Cooper pairs, followed by the development of 
coherence among pairs in the
system forming a condensate~\cite{emery,dagotto}. In this last subsection, we show computational
evidence compatible with pairing of holes in some narrow regions of doping. We certainly do not exclude
that instead of a coherent superconductor, we instead have a pair-density wave. But, as already explained,
recent neutron scattering results for two-dimensional cuprates are compatible with the presence of a pair-density in ladders in the stripe regime, followed by phase
coherence induced by the coupling among ladders~\cite{inter3}. Either way, we believe our results are evidence of
superconductivity tendencies in our models of iron-based ladders.

First let us analyze the pair-correlations for 
various values of hole dopings. We calculate the pair-pair correlations 
at a fixed value of the interaction strength $U/W=3.0$, 
and for cluster size $L=2 \times 12$. 
For our model with two-leg ladder geometry and 
two orbitals ($a$ and $b$) at each site, we could have a variety of pair operators to use in
the pair-pair correlations. Early investigations suggest that properly selecting the pair operator
enhances the signal for superconductivity. However, that task is complicated. For simplicity, 
here we  have focused on intraorbital nearest-neighbor 
sites pairing-operators (certainly, the presence of local on-site Hubbard repulsion
$U$ and $U'$ renders the possibility of electron on-site pairing negligible, opposite to the
negative $U$ Hubbard model)~\cite{patel1}. 
As discussed in previous subsections, the holes 
reside mainly on orbital $a$ and form pairs along the rungs and digonals of the ladder. For this reason,
we calculate the correlation functions focusing only on 
orbital $a$ and along both the rung and plaquette diagonal directions of the ladder. 
We have also checked the pair-correlations along the leg directions,
but we found they decay at a much faster rate compared to all other correlations. To avoid colliding
with one another, the optimal case for the holes in the pair is to have one hole at one leg,
and the other hole at the other leg. 

As a consequence, in order to examine the nature of the dominating superconducting correlation, 
we analyzed the singlet and triplet pair 
correlation functions~\cite{ohta,riera} along the 
vertical and diagonal rungs (see schematic shown in Fig.~4). 
The nearest-neighbor singlet pairing operator for
orbital $a$ along the vertical rung is defined as
\begin{equation}
	O^{\dagger}_{rung,S}(i,a)= \frac{1}{\sqrt{2}}\big[ c_{i,a,1,\ups}^{\dagger} c_{i,a,2,\downs}^{\dagger} -c_{i,a,1,\downs}^{\dagger} c_{i,a,2,\ups}^{\dagger} \big],
\end{equation}
where $i$ is the site index, while 1 and 2 are the leg index. The nearest-neighbor triplet
pairing operator for orbital $a$ along the vertical rung is 
\begin{equation}
	O^{\dagger}_{rung,T}(i,a)= \frac{1}{\sqrt{2}}\big[c_{i,a,1,\ups}^{\dagger} c_{i,a,2,\downs}^{\dagger} +c_{i,a,1,\downs}^{\dagger} c_{i,a,2,\ups}^{\dagger} \big].
\end{equation}
The singlet pairing operator for orbital $a$ along the ladder plaquette diagonal is
\begin{equation}
	O^{ \dagger}_{diag,S}(i,a)= \frac{1}{\sqrt{2}}\big[c_{i,a,2,\ups}^{\dagger} c_{i+1,a,1,\downs}^{\dagger} -c_{i,a,2,\downs}^{\dagger} c_{i+1,a,1,\ups}^{\dagger}\big].
\end{equation}
Finally, the intraorbital spin triplet pairing operator along the ladder plaquette diagonal is
\begin{equation}
	O^{\dagger}_{diag,T}(i,a)= \frac{1}{\sqrt{2}}\big[c_{i,a,2,\ups}^{\dagger} c_{i+1,a,1,\downs}^{\dagger} +c_{i,a,2,\downs}^{\dagger} c_{i+1,a,1,\ups}^{\dagger} \big].
\end{equation}
Using these pairing operators, we have calculate the averaged superconducting
pair-pair correlations for orbital $a$ defined as
\begin{equation}
	P_{a}(d)=\frac{1}{N_d}\sum_{i}\langle O^{\dagger}_{a}(i) O_{a}(i+d) \rangle, 
\end{equation}
where $d=|i-j|$ is the distance along the leg of the ladder, and $N_d$ is the number of equal-distant pairs from site $i$. In other words, we consider all the distances within our clusters, properly normalized. 

\begin{figure}[h]
\centering
\rotatebox{0}{\includegraphics*[width=\linewidth]{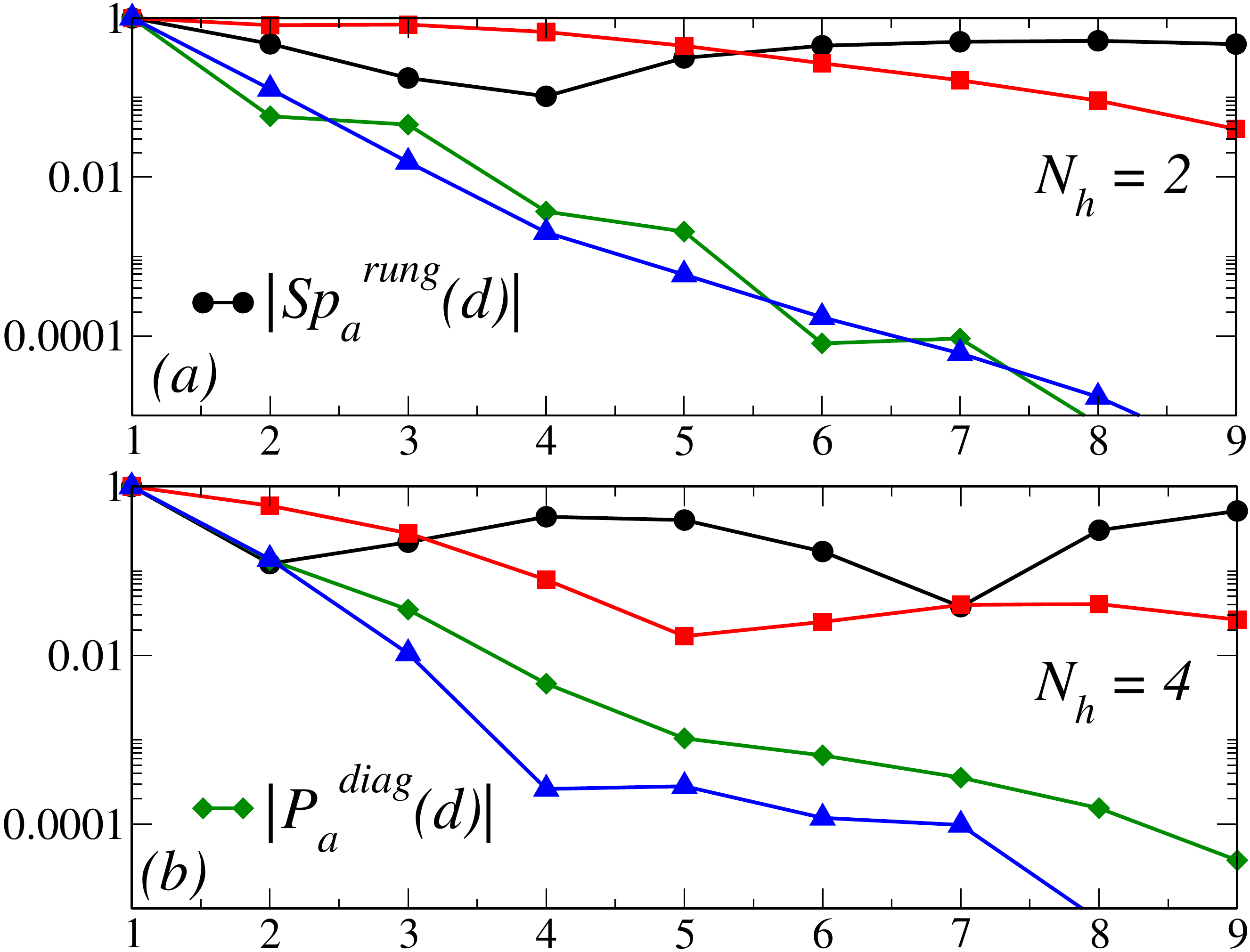}}
\rotatebox{0}{\includegraphics*[width=\linewidth]{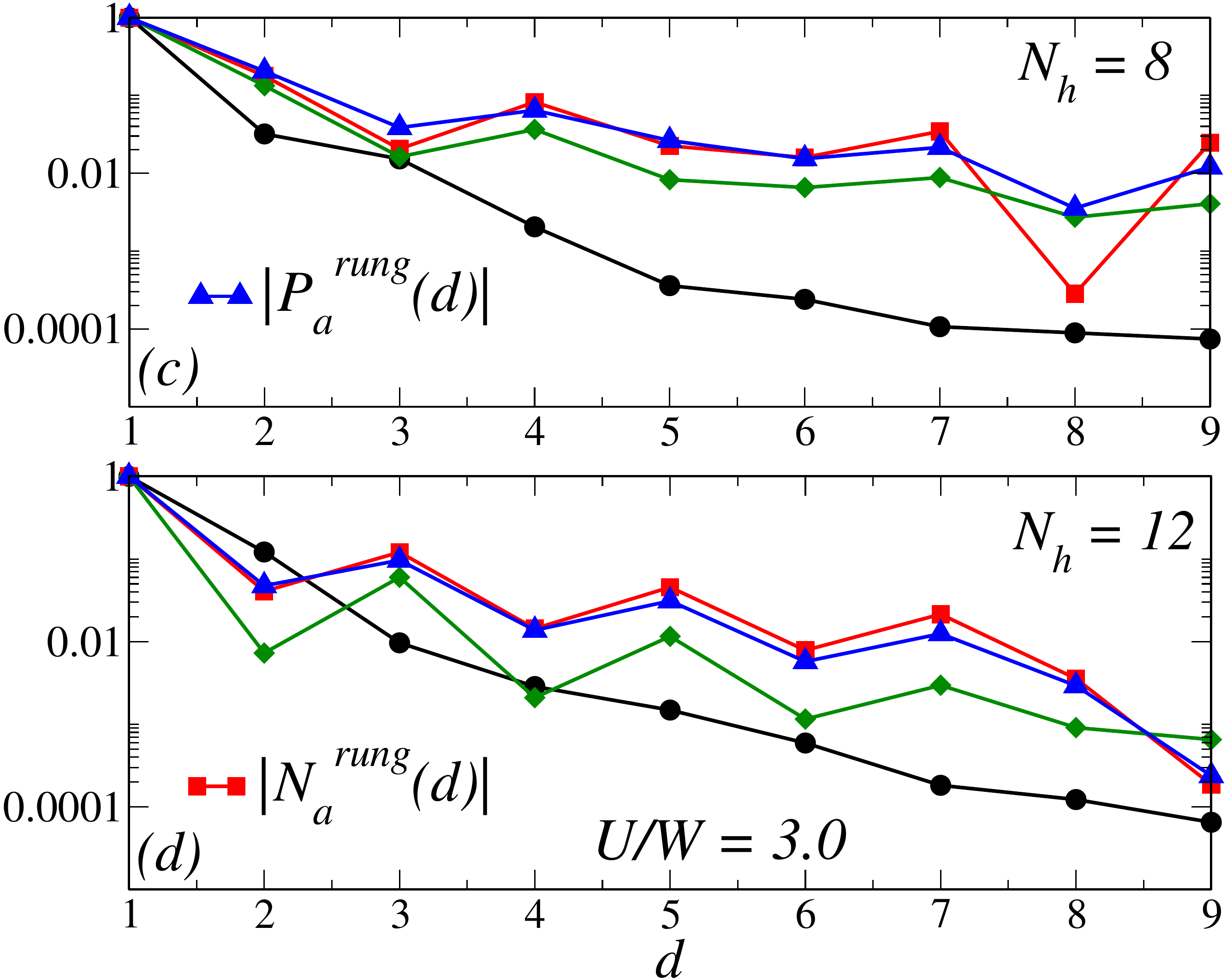}}
	\caption{Comparison of spin-spin  $Sp_a^{rung}(d)$,  
	charge-charge $N_a^{rung}(d)$, and singlet pair-pair
	$P_a^{rung}(d)$ and $P_a^{diag}(d)$ correlations. The various panels correspond to 
	(a) $N_h=2$, (b) $N_h=4$, (c) $N_h=8$, and (d) $N_h=12$ holes, all
	at $U/W=3.0$ and $J_H/U=0.25$ and using a system size $L=2 \times 12$. 
	}
\label{Fig8}
\end{figure}

In Fig.~\ref{Fig7}, we show the spin-singlet pair correlations 
[$P_a^{rung}(S)$ and $P_a^{diag}(S)$] and 
triplet pair correlations [$P_a^{rung}(T)$ and $P_a^{diag}(T)$]
for orbital $a$ with $N_h= 8$ holes. 
While calculating the pair correlations, 
we have discarded the first and last rungs, 
to reduce boundary effects.
We find that the spin singlet correlation along the rung 
$P_a^{rung}(S)$ is the most dominating pair correlation.
This is to be expected  due to that fact that 
holes form a bounded pair along the vertical rung with the highest probability.
The spin-singlet correlation along the diagonal rung $P_a^{diag}(S)$ 
also decays similarly, although with slightly less amplitude, becoming the 
second most dominating pair-pair correlation. 
On the other hand, the spin-triplet pair correlations 
$P_a^{rung}(T)$ and $P_a^{diag}(T)$ decay exponentially (i.e. approximately linearly in the logarithmic scale used).
These numerical results show that the pair-correlation is robust in the
spin-singlet channel and primarily along the rungs of the ladder.

In a quasi-one-dimensional system, 
the slowest decaying correlation functions at long
distance determine the dominant type of ordering. It is expected that once inter chain or ladder
couplings are turned on,
this slowest decaying channel will become dominant. 
In order to compare the pair-pair correlations with spin and charge correlations
for orbital $a$, we have calculated the rung-spin  and rung-charge 
correlations~\cite{nocera}. 
We define the combined rung spin as ${\bf{S}}_{a,i}=  \left({\bf{S}}_{a,i,1} + {\bf{S}}_{a,i,2} \right)$, where 1 and 2 refer to the leg index of the ladder.
The averaged run spin-correlation then becomes 
$Sp_a(d)= \frac{1}{N_{d}} \sum_i \langle {\bf{S}}_{a,i} \cdot {\bf{S}}_{a,i+d} \rangle$. For the rung charge we use the definition $N_{a,i}= \left(N_{a,i,1} + N_{a,i,2} \right)$, and the charge-charge correlations for orbital $a$ are calculated via
$N_{a}(d) =  \frac{1}{N_{d}} \sum_i \langle N_{a,i} N_{a,i+d} \rangle - \langle N_{a,i} \rangle  \langle N_{a,i+d} \rangle$. For spin and charge correlations
also we have discarded the first and last rungs of the ladder, to reduce
boundary effects.

Figure~\ref{Fig8} contains the comparison of pair-pair correlations 
vs. rung-spin and rung-charge correlation, for various values of 
hole dopings, at $U/W=3.0$. For low doping
$N_h=2$ and $N_h=4$, the spin correlation 
$Sp_a^{rung}(d)$ clearly dominates over all other 
correlations [see Fig.~\ref{Fig8}(a) and Fig.~\ref{Fig8}(b)]. This is understandable:
even with pair formation, the spin channel has not been fully scrambled and coherence over
long distances remains. A typical high-$T_c$ superconductor remains antiferromagnetic over
a finite range of doping starting in the undoped parent compound.
For this reason, the singlet pair correlations $P_a^{rung}(d)$ and $P_a^{diag}(d)$ 
decay with much faster rate for lower dopings.

Interestingly, when we increase the hole doping to
$N_h=8$ and $10$ (the latter not shown), 
we find that the rung-singlet pair correlation 
$P_a^{rung}(d)$ starts competing with the charge correlation $N_a^{rung}(d)$.
At large doping, the spin-correlation $Sp_a^{rung}(d)$ 
decays with faster rate due to scrambling of spin-ordering 
in the presence of holes [see Fig.~\ref{Fig8}(c) and (d)].
With further increase in the hole concentration the rung-charge correlation
$N_a^{rung}(d)$ dominates over all other correlations and the
singlet correlation also decreases in magnitude [Fig.~\ref{Fig8}(d)].

Remarkably, as shown in Fig.~\ref{Fig8}{c}, the rung-singlet pair 
correlation $P_a^{rung}(d)$ decays with similar rate as the
rung-charge correlation $N_a^{rung}(d)$ 
for $N_h=8$ holes. A similar decay rate of rung-singlet pair
correlation with charge correlations suggest the existence of  a
pair-density wave state at $U/W=3.0$~\cite{inter1,inter2,inter3}. 
The pair-density wave is an exotic state, with interwined charge density
and superconducting orders. 
We believe superconductivity can be
achieved by weak inter-ladder Josephson coupling~\cite{inter3}, which 
can transform the one-dimensional pair-density-wave state to a two-dimensional coherent
superconducting state. 
In a recent experiment it has been shown
that the superconductivity found in BaFe$_2$S$_3$ has a bulk origin~\cite{nambu}. 
Also they found that increasing pressure
leads to an increase in the inter-ladder transfer of Fe $3d$ electrons, and as a result
 the system turns more metallic~\cite{nambu}. 
 In the case of an array of Cu-based two-leg doped ladders, namely with one active orbital,
 it was found that with intra-ladder coupling 
 leads to a superconducting state~\cite{fradkin,srwhite}.

\subsection{C. Effect of interaction and Hund's coupling }

\begin{figure}[h]
\centering
\rotatebox{0}{\includegraphics*[width=\linewidth]{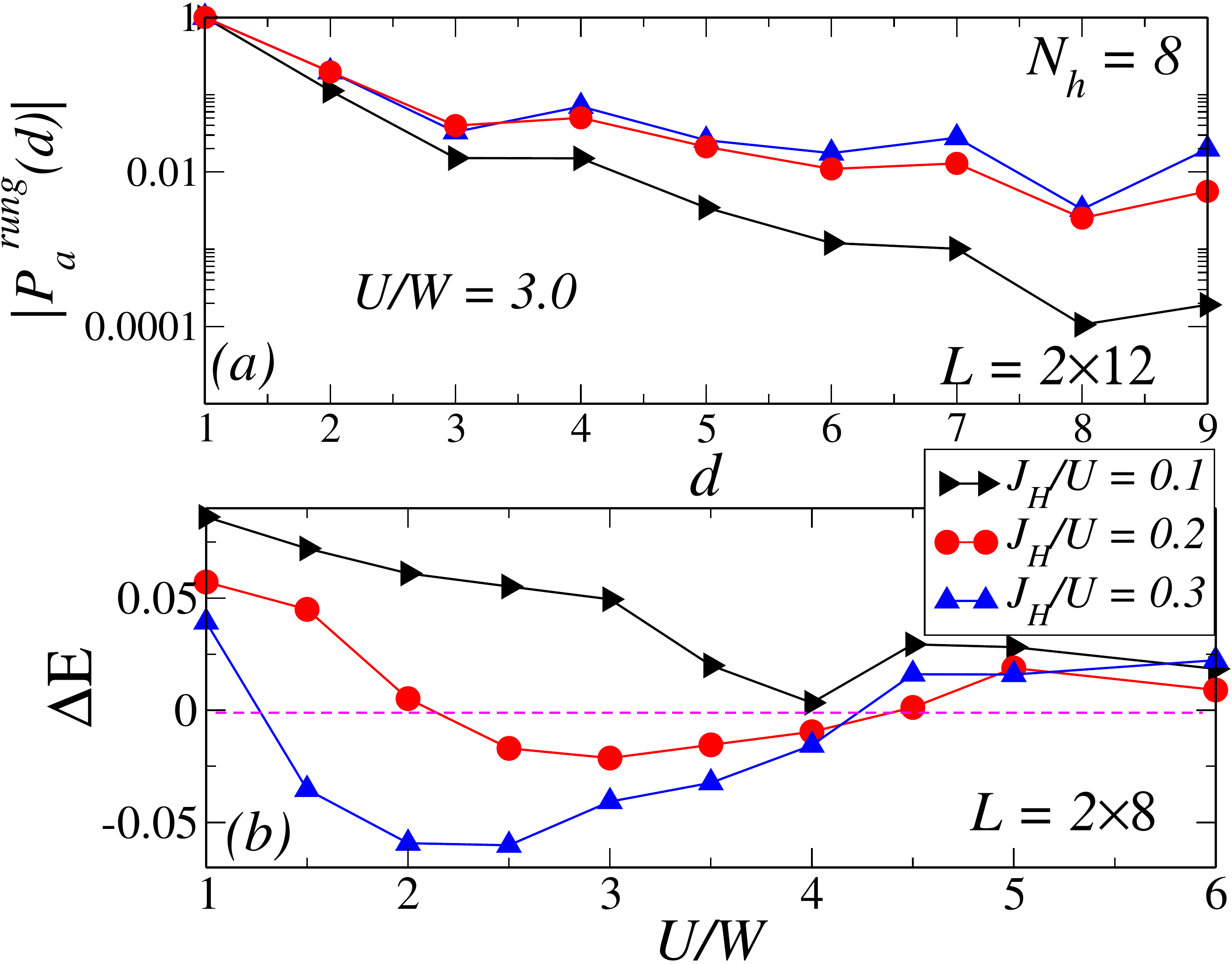}}
	\caption{(a) Rung singlet pair-pair $P_a^{rung}(d)$ at $U/W=3.0$ 
	with $N_h=8$ 
	holes, for different values of $J_H/U$.
	(b) Binding energy $\Delta E$ vs. interaction strength $U/W$
	for different values of Hund's coupling $J_H/U$,
	using a cluster size $L= 2\times 8$.
	}
\label{Fig9}
\end{figure}
In Fig.~\ref{Fig9}(a), we show a comparison of the rung-singlet pair
 correlation $P_a^{rung}(d)$ at $U/W=3.0$ and for three different values 
 of Hund's coupling $J_H/U$, with $N_h=8$ holes. 
 As shown in Fig.~\ref{Fig9}(a), the singlet pair correlation $P_a^{rung}(d)$
 decays with a faster rate at $J_H/U=0.1$. On the other hand, 
the singlet pair correlation $P_a^{rung}(d)$ is enhanced by increasing 
 $J_H/U$. The increase in magnitude of $P_a^{rung}(d)$ is 
 clearly consistent with the results for the binding energy (i.e. $\Delta E$
 become negative for $J_H/U \gtrsim 0.15$).

 We have also studied the effect of Hund's coupling $J_H/U$ on the
bindings of holes. In Fig.~\ref{Fig9}(b), we display the binding energy $\Delta E$
vs. interaction strength $U/W$ for three different values of $J_H/U$.
At $J_H/U=0.1$ binding energy remain positive or close to zero 
for all values of $U/W$, indicating no binding occurs for lower
values of Hund's coupling. However,  for $J_H/U=0.2$ and $0.3$ the
binding energy becomes negative in a wide range of $U/W$. 
The (negative) value of $\Delta E$ increases as $J_H/U$ increases. 
These results clearly show that the
Hund's coupling $J_H/U$ plays an important role
in inducing pairing and in the enhancement of singlet-pair correlations. 

\subsection{D. Pairing tendencies with an alternative set of hopping parameters}
\begin{figure}[h]
\centering
\rotatebox{0}{\includegraphics*[width=\linewidth]{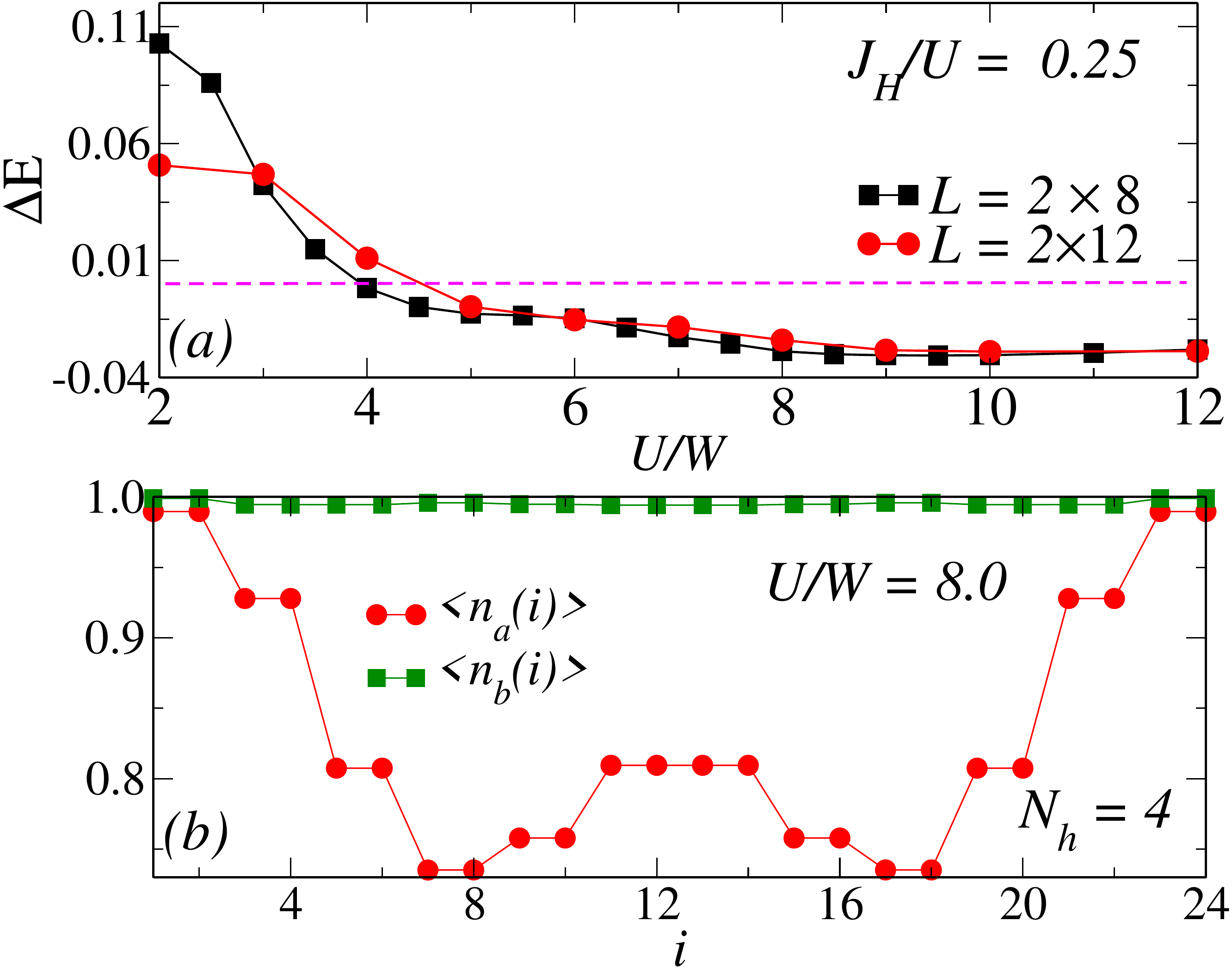}}
	\caption{Results using an alternative set of hoppings.
(a) Binding energy $\Delta E$ vs. interaction strength $U/W$
	at $J_H/U=0.25$ for cluster sizes $L=2 \times 8$ and $2 \times 12$. 
	(b) Real-space charge density $\langle n_{\alpha} \rangle$ vs. site index $i$ with $N_h=4$ holes and at $U/W=8.0$. 
	}
\label{Fig10}
\end{figure}

Let us examine now the binding of holes using an alternative
set of hopping parameters. The new set of hopping parameters at pressure
 $P=12.36$ GPa was derived also using first-principles density functional 
 theory band structure calculations [based on the Vienna {\it ab initio} simulation package (VASP) code~\cite{joubert,bloch,perdew} and the maximally localized Wannier function as implemented in the WANNIER90 code~\cite{marzari,mosto}]. To obtain the hopping matrices,
 we only focus on the band around the Fermi level and considered the hopping matrix with distance up to $\sqrt{2}$ Fe-Fe.
 Thus, modifications on the hopping matrix were performed accordingly to render the 
tight binding band structure to fit  better the {\it ab initio} results.
 It must be clarified that the technical need to constraint the
matching between band structure and tight-binding results using only two orbitals (for practical
reasons, because a DMRG study of a two-leg three-orbital ladder would be too costly)
renders the set of hoppings not unique. Thus, it is convenient to analyze what physics
is obtained with alternative hopping sets. 

Thus, here we will repeat part of the previous results using 
the following 2$\times$2 hopping matrix between sites $i$ and $i+{\hat x}$ and
along the legs of the ladder $t^{x}_{\gamma, \gamma'}$ defined by (in eV units):
\[
t^{x}_{\gamma, \gamma'}=
  \begin{bmatrix}
    -0.65 & +0.25  \\
    -0.25 & +0.216  
  \end{bmatrix}
\]
where $\gamma$ is the orbital index for site $i$ and  $\gamma'$ for $i+{\hat x}$, as before. $t^{y}_{\gamma, \gamma'}$ is the 2$\times$2 hopping matrix along the vertical rung direction:
\[
t^{y}_{\gamma, \gamma'}=
  \begin{bmatrix}
    -0.10 &  0.00 \\
     0.00 & +0.181 
  \end{bmatrix}
\]
The hoppings $t^{x+y}$ and $t^{x-y}$ are 2$\times$2 hopping matrices along the plaquette diagonals of the ladder:
\[
t^{x+y}_{\gamma, \gamma'}= t^{x-y}_{\gamma, \gamma'}=
  \begin{bmatrix}
    +0.05 & -0.322 \\
    +0.322 & +0.084 
  \end{bmatrix}
\]
The crystal fields $\Delta_{\gamma}$  at $P=12.36$~GPa for each orbital are in this case $\Delta_{a}=0.70$~eV and $\Delta_{b}=-0.297$~eV. The kinetic energy  bandwidth is $W=3.238$~eV.

Figure~\ref{Fig10}(a) shows the binding energy $\Delta E$ vs. interaction
strength $U/W$, using this new hopping parameters for cluster sizes 
$L=2 \times 8$ and $L=2 \times 12$ at $J_H/U=0.25$. With the new set of hoppings the 
binding energy takes negative values at larger values of $U/W$,
compared to the older hopping parameters. Both sets of hoppings lead to a 
similar magnitude of $\Delta  E \approx -0.04$~eV in the binding region, in spite of the
different $U/W$ range for the binding.  Clearly, pairing of holes is possible for both set of hopping parameters.

The binding of holes is further confirmed by calculating the
real-space density $\langle n_a(i) \rangle$ for different number of 
holes $N_h=4,6,8$, where we find 2 minima for 4 holes, 3 minima for
6 holes, and 4 minima for 8 holes. Figure~\ref{Fig10}(b) shows the
real-space density plots for orbital $a$ and $b$. The density of orbital
$b$ remains close to one, while there are two minima in the density 
$\langle n_a(i) \rangle$ of orbital $a$, showing the formation of two hole pairs
at $U/W=8.0$. The pairing of holes along the rung and negative values of
binding energy, namely the qualitative similarities with the other set, 
suggest that the new hoppings will likely also develop a pair density wave or directly a superconductor
adding couplings among ladders.

\section{V. Conclusions}
In this publication, we have studied the doped two-orbital Hubbard model
for two-leg ladders with hopping parameters corresponding to the compound
BaFe$_2$S$_3$. Using DMRG calculations for cluster sizes up to $L=2 \times 12$, 
we have investigated magnetic and pairing tendencies for various 
doping strengths. We find different type of magnetic ordering 
 when increasing the interaction strength $U/W$ and hole concentrations.
 For moderate values of interaction strength, the system shows incommensurate
 magnetic ordering, involving short range AFM-correlation along the leg
 and FM-correlation along the rung direction, as in the compound at ambient pressure that in our case
corresponds to the undoped limit. 

Remarkably, in the presence of holes and with increase in $U/W$, the AFM-order along the 
leg direction changes to
 FM-ordering, whereas FM-order along rungs changes to AFM-order. For even
 larger values of $U/W$, FM-order appears along both leg and rung directions.
 Interestingly, we find a robust evidence of pairing of holes. The 
  binding energy becomes negative (pair formation) for wide ranges of interaction 
 parameter $U/W$. The real-space density profile  
 of orbital $a$ shows clear indication of pairing of 
 holes along the rungs of the ladder. Moreover, we have presented a
 comparison of rung-spin, rung-charge, and pair-pair correlations.
 At lower doping concentration, we find the rung-spin (magnetic) correlation dominates
 over other correlations. For moderate values of doping of holes
 we find an exotic pair-density wave state, where the rung-singlet pair correlation and rung-charge correlation coexist. Furthermore, we have shown that the Hund's coupling
 plays an important role in the pairing of holes. In addition, we have also
 shown that the qualitative aspects of the pairing of holes are similar employing 
two sets of hopping matrices, suggesting that our study is generic and survives in a
wide range in parameter space.

\section{Acknowledgments}
The work of B.P, R. S, L.-F.L., and E.D. was supported by the U.S. 
Department of Energy (DOE), Office of Science, Basic Energy Sciences
(BES), Materials Sciences and Engineering Division. 
G.A. was partially supported by the Center for Nanophase Materials Sciences, 
which is a U.S. DOE Office of Science User Facility, and by the Scientific Discovery  through  Advanced  Computing  (SciDAC) program  funded  by  U.S.  DOE,  Office  of  Science, Advanced  Scientific  Computing  Research  and  Basic Energy Sciences, Division of Materials Sciences and Engineering.
Validation and some computer runs were conducted at the Center for Nanophase Materials Sciences, which is a DOE Office of Science User Facility.

\end{document}